\pdfoutput=1
\documentclass[11pt]{article}
\usepackage[hmargin=2.54cm,vmargin=2.54cm]{geometry} % see geometry.pdf
\usepackage{geometry}
\usepackage{amsmath,amssymb,setspace,fancyhdr,geometry,url,color}
\usepackage{subfigure,graphicx,caption}
\usepackage{lscape,amsthm}
\usepackage[authoryear,round]{natbib}
\usepackage{tabularx}
\RequirePackage{lineno} 
\allowdisplaybreaks[0]
\allowbreak
\title{Calibrating an Ice Sheet Model using High-dimensional Binary Spatial Data}
\author{Won Chang, Murali Haran, Patrick Applegate, David Pollard}
\begin{document}
%\linenumbers
\newcommand{\Dir}{\mathrm{Dir}}
\newcommand{\ceil}[1]{\lceil #1 \rceil}
\newcommand{\thh}{^\mathrm{th}}
\newcommand{\modtwo}{\mathrm{[mod~2]}}
\newcommand{\thetaof}[2]{\theta \langle #1;#2\rangle}
\newcommand{\Mpa}{M_\mathrm{P,A}}
\newcommand{\Ma}{M_\mathrm{A}}
\newcommand{\rjaccept}{\mathcal{A}}

\newcommand{\matern}{Mat\'{e}rn }
\newcommand{\ba}{\ensuremath{\mathbf{a}}}
\newcommand{\bA}{\ensuremath{\mathbf{A}}}
\newcommand{\balpha}{\ensuremath{\boldsymbol{\alpha}}}
\newcommand{\barB}{\ensuremath{\mathbf{\bar{B}}}}
\newcommand{\barY}{\ensuremath{\bar{\mathbf{Y}}}}
\newcommand{\barZ}{\ensuremath{\mathbf{\bar{Z}}}}
\newcommand{\bb}{\ensuremath{\mathbf{b}}}
\newcommand{\bB}{\ensuremath{\mathbf{B}}}
\newcommand{\bC}{\ensuremath{\mathbf{C}}}
\newcommand{\bD}{\ensuremath{\mathbf{D}}}
\newcommand{\bV}{\ensuremath{\mathbf{V}}}
\newcommand{\bdelta}{\ensuremath{\boldsymbol{\delta}}}
\newcommand{\be}{\ensuremath{\mathbf{e}}}
\newcommand{\bepsilon}{\ensuremath{\boldsymbol{\epsilon}}}
\newcommand{\bG}{\ensuremath{\mathbf{G}}}
\newcommand{\bIn}{\ensuremath{\mathbf{I}_n}}
\newcommand{\bIN}{\ensuremath{\mathbf{I}_N}}
\newcommand{\bJ}{\ensuremath{\mathbf{J}}}
\newcommand{\bk}{\ensuremath{\mathbf{k}}}
\newcommand{\bK}{\ensuremath{\mathbf{K}}}
\newcommand{\bLambda}{\ensuremath{\boldsymbol{\Lambda}}}
\newcommand{\bM}{\ensuremath{\mathbf{M}}}
\newcommand{\bmu}{\ensuremath{\boldsymbol{\mu}}}
\newcommand{\bnu}{\boldsymbol{\nu}}
\newcommand{\bolda}{\ensuremath{\mathbf{a}}}
\newcommand{\boldeta}{\ensuremath{\boldsymbol{\eta}}}
\newcommand{\bomega}{\ensuremath{\boldsymbol{\omega}}}
\newcommand{\bP}{\ensuremath{\mathbf{P}}}
\newcommand{\bphi}{\ensuremath{\boldsymbol{\phi}}}
\newcommand{\bpsi}{\ensuremath{\boldsymbol{\psi}}}
\newcommand{\bQ}{\ensuremath{\mathbf{Q}}}
\newcommand{\bR}{\ensuremath{\mathbf{R}}}
\newcommand{\bs}{\ensuremath{\mathbf{s}}}
\newcommand{\bSigma}{\ensuremath{\boldsymbol{\Sigma}}}
\newcommand{\btheta}{\ensuremath{\boldsymbol{\theta}}}
\newcommand{\bu}{\ensuremath{\mathbf{u}}}
\newcommand{\bh}{\ensuremath{\mathbf{h}}}
\newcommand{\bx}{\ensuremath{\mathbf{x}}}
\newcommand{\bv}{\ensuremath{\mathbf{v}}}
\newcommand{\bw}{\ensuremath{\mathbf{w}}}
\newcommand{\bW}{\ensuremath{\mathbf{W}}}
\newcommand{\bX}{\ensuremath{\mathbf{X}}}
\newcommand{\bxi}{\ensuremath{\boldsymbol{\xi}}}
\newcommand{\bY}{\ensuremath{\mathbf{Y}}}
\newcommand{\bz}{\ensuremath{\mathbf{z}}}
\newcommand{\bZ}{\ensuremath{\mathbf{Z}}}
\newcommand{\bzero}{\ensuremath{\mathbf{0}}}
\newcommand{\cl}{\ensuremath{c\ell}}
\newcommand{\Cov}{\ensuremath{\mbox{Cov}}}
\newcommand{\E}{\ensuremath{\mbox{E}}}
\newcommand{\Eta}{\ensuremath{\mbox{H}}}
\newcommand{\sigep}{\sigma_{\boldsymbol{\epsilon}}^2}
\newcommand{\Sigom}{\Sigma_{\boldsymbol{\omega}}}
\newcommand{\bU}{\ensuremath{\mathbf{U}}}
\newcommand{\pkg}[1]{{\normalfont\fontseries{b}\selectfont #1}}
\let\proglang=\textsf
\let\code=\texttt
\renewcommand{\thefootnote}{\fnsymbol{footnote}}

\maketitle
\doublespacing

\begin{abstract}

Rapid retreat of ice in the Amundsen Sea sector of West Antarctica may cause drastic sea level rise, posing significant risks to populations in low-lying coastal regions. Calibration of computer models representing the behavior of the West Antarctic Ice Sheet is key for informative projections of future sea level rise. However, both the relevant observations and the model output are high-dimensional binary spatial data; existing computer model  calibration methods are unable to handle such data. Here we present a novel calibration method for computer models whose output is in the form of binary spatial data. To mitigate the computational and inferential challenges posed by our approach, we apply a generalized principal component based dimension reduction method. To demonstrate the utility of our method, we calibrate the PSU3D-ICE model by comparing the output from a 499-member perturbed-parameter ensemble with observations from the Amundsen Sea sector of the ice sheet. Our methods help rigorously characterize the parameter uncertainty even in the presence of systematic data-model discrepancies and dependence in the errors. Our method also helps inform environmental risk analyses by contributing to improved projections of sea level rise from the ice sheets.

\end{abstract}

\vspace*{.2in}


\section{Introduction}

Mass loss from the polar ice sheets has the potential to make substantial contributions to future sea level rise. The ice sheets therefore pose substantial risks to people living near present sea level. Taken together, the ice sheets contain sufficient ice to raise global mean sea level by up to 60 meters if they were to melt completely; the West Antarctic Ice Sheet (WAIS) might contribute up to 4 meters to future sea level rise \citep{fretwell2013bedmap2}, whereas the Greenland and East Antarctic Ice Sheets might contribute up to 7 meters and up to 50 meters to sea level rise, respectively. Recent observations of increased surface air temperatures at high northern latitudes, and warmer ocean waters at high southern latitudes, suggest that some fraction of this ice may melt over the next few decades or centuries \citep{zhang2007increasing,serreze2011processes}.  Geologic observations also indicate that the ice sheets can make rapid contributions to sea level \citep[e.g.,][]{deschamps2012ice}.  A significant fraction of the world's population lives near present sea level \citep{nicholls2008global}, and these people will face an increased risk from storm surges as sea level rise progresses.  The future behavior of the ice sheets is hence a key input for risk analyses \citep[e.g.,][]{lempert2012characterizing}.

Despite the ice sheets' importance for future sea level rise, projecting their future behavior using computer models remains challenging.  In contrast to global climate models, which represent the behavior of the atmosphere, oceans, vegetation, and other parts of the Earth system, ice sheet models typically include advanced treatments of ice flow and simplified representations of the boundary conditions at the ice sheets' interfaces with the atmosphere and oceans.  Until recently, many ice sheet models lacked realistic representations of processes that may play an important role in the ice sheets' future development \citep[e.g.,][]{little2007towards}.  Recent studies \citep[see, e.g.][]{stone2010investigating,applegate2012assessment} confirm that imperfect knowledge of model input parameters leads to large uncertainties in modeled sea level contributions.  Thus, characterizing and reducing parametric uncertainty is a key step in producing ice sheet projections that can properly inform climate risk analyses.

Calibration of ice sheet models is challenging because the relevant data are high-dimensional binary spatial data, and the models themselves are computationally expensive.  Recent efforts to calibrate ice sheet models include \citet{gladstone2012calibrated}, \citet{mcneall2013potential}, and \citet{chang2013ice}. These studies represent significant advances in the state of the art for ice sheet model calibration, but each has significant limitations. 
\citet{gladstone2012calibrated} uses a heuristic approach which chooses input parameter settings that yield output values which fall in the 95\% confidence set. The confidence set is defined by the  `three-sigma rule' \citep{pukelsheim1994three}, choosing the margin of errors for the confidence sets as $\pm 3~ \mbox{standard errors}$, and a Bonferroni-type multiple test procedure to generalize it to the multivariate case. The approach ensures robustness against assumptions about the model-observation discrepancy and the observational errors. The framework requires running the ice sheet model for at least thousands of parameter settings and hence is applicable only to heavily simplified ice sheet models such as the one-dimensional flow-line model used by \citet{gladstone2012calibrated}. \citet{mcneall2013potential} uses a Gaussian process emulator for three quantities: the overall ice volume, the maximum ice height, and the area covered by ice. The use of Gaussian process emulators makes this approach computationally  feasible, even with computationally expensive 3-D ice-sheet models. However, their analysis is based on heavily aggregated quantities and therefore does not utilize spatially resolved information. Moreover, instead of utilizing the probability model given by the emulator, they rely on a heuristic approach based on an ``implausibility measure'', which is the model output mean standardized using the observational data. The approach in \citet{chang2013ice} improves upon these two approaches by using the probabilistic calibration approach proposed by \citet{kennedy2001bayesian} and extended and generalized by many others \citep[see, e.g.][]{kennedy2001bayesian,bayarri2007computer,Higdon2008}. 

However, current calibration approaches are unable to utilize important sources of information about ice sheet behavior such as the extent of ice covered area in modern- and paleo-observations \citep{alley2010history} due to the binary nature of the model output and observations. For example, some Greenland ice cores contain ice from the last interglacial period (from 125,000 years ago), whereas others do not; however, rigorous calibration approaches that can take this type of data into account are lacking. \citet{chang2013ice} circumvents this problem by using datasets obtained by aggregating the 2-dimensional spatial pattern of ice thickness into a 1-dimensional profile to make the data more suitable for Gaussian process-based approaches and hence is silent on using datasets that are in the form of non-Gaussian spatial processes. Aggregating data in this manner likely results in significantly larger projection uncertainties than would result if unaggregated data were used \citep{chang2013fast}. Thus, more work to properly incorporate binary spatial data into ice sheet model calibration is needed.

Here we propose a new calibration approach for high-dimensional binary spatial data, appropriate for calibration of ice sheet models using data on the spatial pattern of ice coverage. The spatial data that we use here are gridded binary patterns, where the outcome at each location represents absence or presence of grounded ice. Our novel calibration approach extends the Gaussian process-based calibration framework to high-dimensional binary spatial data. Our approach is based on a generalized linear model framework with a latent Gaussian process. To mitigate the considerable computational challenges posed by the high-dimensional latent process, we reduce the dimensionality of the model output and the data using a logistic principal component analysis \citep{collins2001generalization, lee2010sparse}. This approach avoids inference with high-dimensional latent variables while also sidestepping the computational burden posed by high-dimensional matrix computations.

We use this approach to calibrate an innovative ice sheet model (PSU3D-ICE; see below), using data on the spatial extent of grounded ice (i.e., binary pattern of presence and absence of grounded ice) within the Amundsen Sea sector of the West Antarctic Ice Sheet.  
The Amundsen Sea sector is likely the largest contributor to sea level rise of any Antarctic ice drainage  \citep{pritchard2012antarctic}. It includes the rapidly thinning Pine Island and Thwaites Glaciers, which drain ice from the interior of the West Antarctic Ice Sheet into the ocean.  The glaciers terminate in ice shelves where the grounded ice becomes too thin to remain in contact with the seafloor.  Ice velocities are high within the glacier trunks, but slow elsewhere.  The position of the grounding line, or the narrow region separating the grounded ice from the floating ice shelves, is known from modern geophysical measurements \citep{fretwell2013bedmap2}.   

The Pennsylvania State University three-dimensional ice sheet model, PSU3D-ICE, simulates the combined ice sheet-ice shelf system using a so-called ``hybrid" dynamical core, which smoothly merges the shallow-shelf and shallow-ice approximations to the Stokes equations describing ice flow \citep{kirchner2011capabilities}.  This approach allows the model to simulate the full ice sheet-ice shelf system realistically, while minimizing the computational cost of the model. \citet{pollard2009modelling,pollard2012simple,pollard2012description} provide detailed descriptions of the model, which has been part of several ice-sheet model intercomparisons \citep[e.g., seaRISE,][]{bindschadler2013ice}.

The remainder of this paper is organized as follows. In Section \ref{section:DataDescription} we describe the model output and the observational data that we use in our analysis. In Section \ref{section:BinaryCalibEmul} we introduce the basic framework for calibration using binary spatial data and explain the computational challenges posed by high-dimensional spatial data. We formulate a reduced-dimensional approach to mitigate these challenges in Section \ref{section:RedDimBinaryCalibration}. We describe how to compute the logistic principal components (PCs) for binary computer model output and formulate an emulation framework based on them, and then set up a calibration model that relies on the reduced-dimensional emulator. Section \ref{section:BinaryCalibImplementation} describes the implementation details and results for our experiments on a simulated example as well as the application of our methods to the ice sheet model for WAIS described in Section \ref{section:DataDescription}. We also discuss the preliminary implications of our research on the study of the behavior of the West Antarctic Ice Sheet. We conclude this paper with a summary and future directions in Section \ref{section:BinaryCalibDiscussion}. 

\section{Model Description and Observational Data} \label{section:DataDescription}

We use an innovative 3-D ice-sheet model, which uses boundary-layer techniques to capture challenging grounding-zone dynamics while still being efficient enough to feasibly perform  long-term simulations. Basically the model simulates the long-term evolution of ice extent and thickness, changing due to slow deforming flow and basal sliding under its own weight, surface snowfall and melt, ocean melting under floating ice shelves, and iceberg calving, over the Antarctic continent. For the
simulations in this manuscript, the model is applied to a nested domain
spanning West Antarctica, with a relatively fine grid resolution of 20 km.
This nested grid is driven at its lateral boundaries by stored results
from a previous coarser-scale continental simulation.

The model is run over the last 40,000 years, initialized appropriately at 40 ka (40,000 years 
Before Present or BP, relative to 1950 AD) from a previous long-term run. 
Atmospheric forcing is supplied using a modern climatological 
Antarctic dataset \citep[ALBMAP:][]{lebrocq2010improved}, with uniform cooling perturbations applied 
proportional to a deep-sea-core $\delta^{18}$O record \citep{pollard2009modelling,pollard2012description}. 
Oceanic forcing is supplied using archived ocean temperatures from a coupled 
AOGCM simulation of the last 20 kyr (1 kyr=1,000 years) \citep{liu2009transient}. Sea level versus time, rising during this period primarily due to Northern Hemispheric ice-sheet melt, is prescribed from the  ICE-5G dataset \citep{peltier2004global}. Modern bedrock elevations are obtained from 
the Bedmap2 dataset \citep{fretwell2013bedmap2}. 

After reaching present day (0 ka), each run is extended for 5,000 years with
a very basic representation of future warming. Atmospheric and oceanic
temperatures are uniformly increased by 6 and 2 $^{\circ}$C, respectively,
ramped up linearly from 0 to 150 years after present and held constant
thereafter. Ocean-temperature increases are confined to a longitudinal sector
around the Amundsen Sea Embayment of West Antarctica, corresponding to
the location of observed sub-ice melt increases in recent decades. This simple prescription of future temperature changes causes drastic ice retreat into the West Antarctic interior in many of the runs. More realistic future warming scenarios are planned for future work. Other aspects of the model are described in \citet{pollard2012simple,pollard2012description}.

All 3-D ice sheet models have a number of poorly constrained internal 
parameters. To evaluate the effects of these uncertain parameters in the PSU3D-ICE model, we selected 10 highly uncertain parameters to vary in a Latin hypercube (LHC) ensemble (Table \ref{table:parameters}). LHC design \citep[see, e.g.][]{mckay1979comparison,stein1987large} allows us to explore a reasonably wide range of parameter values with a relatively smaller number of model runs even with a high-dimensional parameter space.  We use a random LHC design generated by `lhc' package in R \citep{carnell2012lhs}. While a random LHC can result in a design with highly correlated input parameter values \citep[cf.][]{joseph2008orthogonal}, we have confirmed that in our design the correlation coefficients between the parameter values are very low (all less than 0.08). Four of these parameters are generally recognized as particularly important, with strong effects on past and present ice configurations, but which represent uncertain processes in glaciology. See Section \ref{subsection:ObsCalibration} for a description of these parameters. Each parameter value is remapped to the $[0,1]$ interval with 0.5 generally being the non-statistically tuned value by previous research \citep{pollard2012description,pollard2015potential} and 499 design points for the input parameters are chosen by LHC design for the 10-dimensional unit cube. 

\begin{figure}
\centering
\includegraphics[scale=0.41]{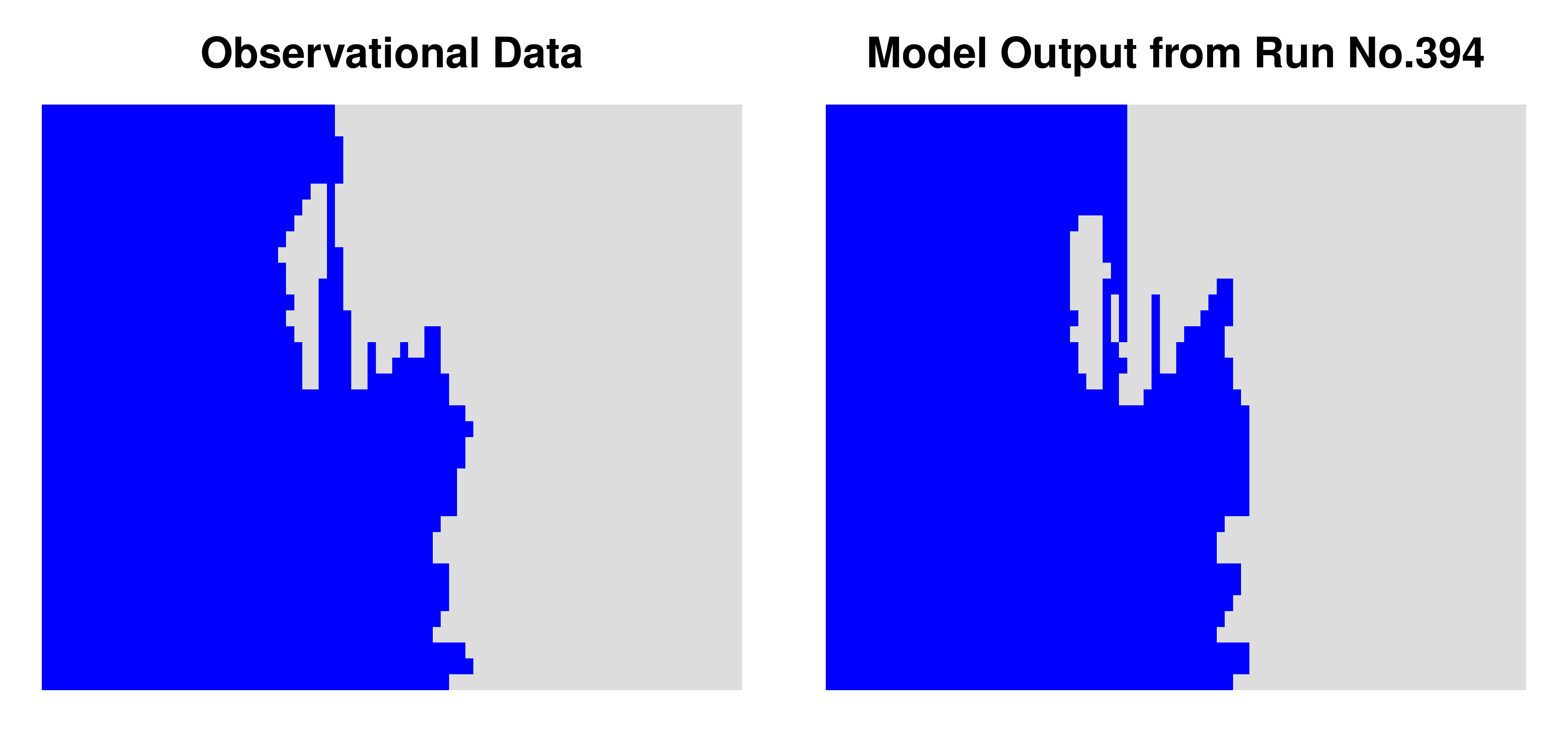}
\caption{Spatial patterns of grounded ice from the observational dataset \citep{fretwell2013bedmap2} (left) and an example PSU3D-ICE run (right). The gray pixels represent ice-covered locations. The  model appears to be capable of reproducing observational patterns.}
\label{fig:ObsAndBestRun}
\end{figure}

We calibrate the model by inferring these ten parameters using the observed modern day grounding-line geometry, i.e., the modern ``coastline'' map of grounded ice versus floating ice, from a modern Antarctic dataset \citep[Bedmap2;][]{fretwell2013bedmap2}. The original model output covers the entire WAIS, but we use a subset of the output that corresponds to the observational data for ASE (Amundsen Sea Embayment). The spatial pattern for each parameter setting and the observational data has $86 \times 37$ regular grid points. At each grid cell, we have a binary outcome with 1 representing presence of grounded ice and 0 representing absence of grounded ice. See Figure \ref{fig:ObsAndBestRun} for the observational data and an example model output. Note that both the model output and the observational data are in the form of high-dimensional and binary spatial patterns; the challenges posed by these data structures will drive much of our methods development in the following sections. 

\section{Computer Model Emulation and Calibration Using Binary Spatial Data} \label{section:BinaryCalibEmul}
In this section, we introduce a framework for computer model emulation and calibration with Gaussian spatial data and then describe our new approach for emulation-calibration with binary spatial data. 
We follow the two-stage approach to computer model calibration \citep{bayarri2007computer,bhat2013inferring,chang2013fast,chang2013composite}.  In the two stage approach, the computer model is first approximated using a Gaussian process emulator, and then in the calibration step, the model parameters are inferred based on observational data using the emulator.  Separating the emulation step from the calibration step makes it easier to diagnose the performance of the emulator and improves the identifiability of the parameters in the calibration model \citep[see][for more discussion]{rougier2008comment,bhat2010computer,bhat2013inferring}. Here the computer model output is in the form of binary spatial data, and we specify a generalized linear model framework and formulate an emulator that interpolates natural parameters across the model input settings using a Gaussian process. For the calibration step we introduce a model for data-model discrepancy and model the natural parameters for the observational data using the constructed emulator and the discrepancy model. When the model output is in the form of Gaussian spatial data, there is a possibility that we underestimate the parametric uncertainty when we do not account for the uncertainty in the estimated emulator parameters \citep[see for instance,][]{bayarri2007computer,bhat2013inferring,liu2009modularization,chang2013fast}. However, when the model output is in the form of binary spatial data, based on several simulated examples we find that we do not underestimate the parametric uncertainty. In fact, allowing uncertainties for this parameters leads to identifiability issues in the inference in the calibration stage.

Computation presents considerable challenges. We therefore propose a principal components-based framework in order to make our approach computationally feasible. To help set the stage for the description of our framework for emulation and calibration in this manuscript, we begin with a description of emulation-calibration using  Gaussian spatial data in Section \ref{section:GaussianCalibEmul}. This framework is also described in \citet{kennedy2001bayesian,bayarri2007computer,Higdon2008,bhat2013inferring,chang2013fast}. Sections \ref{section:BinaryEmul} and \ref{section:BinaryCalib} provide details about our framework for emulation and calibration for binary spatial model output and observational data. Section \ref{section:BinaryChallenges} describes the computational and inferential challenges, and Section \ref{section:RedDimBinaryCalibration} provides details about our computationally expedient reduced-dimension approach to mitigate these challenges.

We use the following notation henceforth.  Let $Y(\btheta,\bs)$ be the computer model output at a parameter setting $\btheta\in \boldsymbol{\Theta}$ and a spatial location $\bs\in\mathcal{S}$ where $\boldsymbol{\Theta}\subset R^d$ is the parameter space and $\mathcal{S}$ is the spatial field that covers the geographical area of interest. The integer $d$ represents the number of input parameters that the computer model has. We assume that we have obtained model runs at $p$ design points $\btheta_1,\dots,\btheta_p\in\boldsymbol{\Theta}$ and each model run has $n$ spatial locations $\bs_1,\dots,\bs_n \in \mathcal{S}$. We let $\bY$ be a $p\times n$ matrix of the computer model output where the $(i,j)$th element $Y(\btheta_i,\bs_j)$ is the model output at the $i$th parameter setting $\btheta_i$ and the $j$th spatial location $\bs_j$. Similarly, let $\bZ$ be the $n$-dimensional observational data where its $j$th element $Z(\bs_j)$ is the observation at the $j$th spatial location $\bs_j$. For the ice sheet model calibration example in Section \ref{subsection:ObsCalibration}, $p=499$ and $n=86\times37=3,182$.

\subsection{Computer Model Emulation and Calibration Using Gaussian Spatial Data} \label{section:GaussianCalibEmul}
We briefly outline emulation  and calibration for Gaussian spatial data  \citep[cf.][]{bhat2010computer,bhat2013inferring,chang2013fast,chang2013composite}. The purpose of this subsection is to make it easier for readers to follow the remainder of this section. Here we assume that $Y(\btheta, \bs)$ can be reasonably modeled by a Gaussian distribution. In the emulation stage, we fit the following Gaussian process model for the computer model output $\bY$ to interpolate the values at the parameter settings $\btheta_1,\dots,\btheta_p$ and the spatial locations $\bs_1,\dots,\bs_n$ simultaneously:
\begin{equation} \label{equation:GaussianEmul}
\mbox{vec}\left( \bY \right) \sim N (\bX \boldsymbol{\beta}, \bSigma (\bxi_y)),
\end{equation}
where $\mbox{vec}(\cdot)$ is the vectorization operator that concatenates columns into one vector. The $np \times b$ matrix $\bX$ contains values of $b$ different covariates that can be used to represent general trends in the model output. The $np\times np$ covariance matrix $\bSigma (\bxi_y)$  with a vector of parameters $\bxi_y$ is defined by a positive definite covariance function for a Gaussian process \citep[see, e.g.][]{sacks1989design} that interpolates between the model output at different parameter settings $\btheta_1,\dots,\btheta_p$ and spatial locations $\bs_1,\dots,\bs_n$. In general, the mean term $\bX \boldsymbol{\beta}$ can be set to $\mathbf{0}$ with a proper centering of $\bY$, and the covariance parameters in $\bxi_y$ can be estimated by the maximum likelihood estimate $\hat{\bxi}_y$.

In the calibration step, we model the observational data using the following model: 
\begin{linenomath}
\begin{equation*}
\bZ = \boldeta (\btheta, \bY) + \bdelta,
\end{equation*}
\end{linenomath}
where $\boldeta(\btheta,\bY)$ is the $n$-dimensional emulated vector, a computer model output approximated by interpolating the model runs in $\bY$ using the model in \eqref{equation:GaussianEmul}, and $\bdelta \sim N(\mathbf{0},\bSigma(\bxi_d))$ is an $n$-dimensional spatially correlated zero-mean Gaussian vector with covariance parameter $\bxi_d$ that represents the systematic model-observation discrepancy. By choosing standard priors for the parameters in the model, one can infer $\btheta$ via Markov Chain Monte Carlo (MCMC) along with other parameters. 

\subsection{Computer Model Emulation Using Binary Spatial Model Output} \label{section:BinaryEmul}

 In this section and for the remainder of this paper, we assume that $Y(\btheta,\bs)$ is binary with 0-1 values.  We assume the model output $Y(\btheta_i,\bs_j)$ to be a Bernoulli random variable with success probability $p_{ij}=P(Y(\btheta_i,\bs_j)=1)$. We denote the corresponding logit $\log \left( \frac{p_{ij}}{1-p_{ij}} \right)$ by $\gamma_{ij}$  and hence the probability mass function for $Y(\btheta_i,\bs_j)$ can be written as

\begin{linenomath}
\begin{equation*} \label{equation:Ybernoulli}
\begin{aligned} 
P(Y(\btheta_i,\bs_j)=y_{ij})=&p_{ij}^{y_{ij}}(1-p_{ij})^{1-y_{ij}}\\
=&\left(\frac{\exp(\gamma_{ij})}{1+\exp(\gamma_{ij})}\right)^{y_{ij}} \left(\frac{1}{1+\exp(\gamma_{ij})}\right)^{1-y_{ij}}\\
=&(1+\exp(-(2y_{ij}-1)\gamma_{ij}))^{-1}.
\end{aligned}
\end{equation*}
\end{linenomath}
We let $\boldsymbol{\Gamma}$ be the $p\times n$ matrix of natural parameters for model output, where its $(i,j)$th element is $\gamma_{ij}$. By assuming that the elements in the model output $\bY$ are conditionally independent given the natural parameters in $\boldsymbol{\Gamma}$, the likelihood function for the model output $\bY$ can be written as follows;
\begin{equation} \label{equation:BinaryLikelihood}
L(\bY|\boldsymbol{\Gamma})=\prod_{i=1}^{p} \prod_{j=1}^{n} P(Y(\btheta_i,\bs_j)=y_{ij}) =\prod_{i=1}^{p} \prod_{j=1}^{n} (1+\exp(-(2y_{ij}-1)\gamma_{ij}))^{-1}.
\end{equation}
The dependence among the elements in $\bY$ is modeled through the dependence among the elements in $\boldsymbol{\Gamma}$. To approximate the computer model we interpolate the natural parameters at different input parameter settings using a Gaussian process \citep[cf.][]{sacks1989design} with zero mean and a covariance function that depends on the input parameter settings and the spatial locations for model output, i.e.,
\begin{equation}\label{eqn:CovGamma}
\Cov\left(\gamma_{ij},\gamma_{kl}\right)=C\left(\btheta_i,\btheta_k,\bs_j,\bs_l;\bxi_y\right),
\end{equation}
where $C$ is a valid covariance function over $(\Theta \times \mathcal{S})^2$ and $\bxi_y$ is a vector of covariance parameters. Note that estimating the natural parameters $\{\gamma_{ij},i=1\dots,p,j=1,\dots,n\}$ is not possible by simply maximizing  \eqref{equation:BinaryLikelihood} since this is an ill-posed problem without any constraints on them (see Section \ref{section:BinaryChallenges} for details).  If estimates of the natural parameters are available, we can fit the Gaussian process to those natural parameters by maximizing the likelihood function corresponding to the probability model 
\begin{linenomath}
\begin{equation*}
\mbox{vec}\left( \boldsymbol{\Gamma} \right) \sim N (\mathbf{0}, \bSigma (\bxi_y))
\end{equation*}
\end{linenomath}
 with respect to $\bxi_y$, where the covariance matrix $\bSigma (\bxi_y)$ is given by a function of the form in \eqref{eqn:CovGamma}. We then use the fitted process to predict the vector of natural parameters at any new parameter value $\btheta$. We call the fitted Gaussian process an ``emulator'' and the resulting predictive process an ``emulator process''. 

\subsection{Computer Model Calibration Using Binary Spatial Data} \label{section:BinaryCalib}
Again, in this section and for the remainder of this paper, we assume that $Z(\bs)$ is binary with 0-1 values. We set up a calibration model that links the input parameter to the observational data using the emulator constructed in the previous section while considering the systematic discrepancy. Let $\boldsymbol{\lambda}=(\lambda_1,\dots,\lambda_n)^T$ be the $n$-dimensional vector of natural parameters for the observational data, then our calibration model is given by
\begin{linenomath}
\begin{equation} \label{eqn:LikelihoodLambda}
\boldsymbol{\lambda}=\boldeta\left(\btheta,\bY\right)+\bdelta,
\end{equation}
\end{linenomath}
where $\bdelta$ is an $n$-dimensional vector for the model-observation discrepancy. Here we can think of $\btheta$ as a ``fitted'' value for the parameter $\btheta$, that is, $\btheta$ at which the values of the natural parameters match those from which the observations are assumed to originate. For $\bdelta$ we use an $n$-dimensional Gaussian random vector with covariance parameters $\bxi_d$ that is independent of the emulator process $\boldeta\left(\btheta,\bY\right)$. This corresponds to the following probability model for each observation given by the standard logistic regression framework:
\begin{linenomath}
\begin{align} \label{eqn:LikelihoodZ}
\begin{split}
P(Z(\bs_j)=z_j)=&\left(\frac{\exp(\lambda_j)}{1+\exp(\lambda_j)}\right)^{z_j} \left(\frac{1}{1+\exp(\lambda_i)}\right)^{1-z_j}\\
=&(1+\exp(-(2z_j-1)\lambda_j))^{-1},
\end{split}
\end{align}
\end{linenomath}
where $\lambda_j$ is the $j$th element of $\boldsymbol{\lambda}$. The log-likelihood function for $\boldsymbol{\lambda}$, $\btheta$, and $\bxi_d$ given $\boldsymbol{\Gamma}$ and $\bZ$ can be written as 
\begin{linenomath}
\begin{align*}
\ell(\bZ,\boldsymbol{\Gamma}|\boldsymbol{\lambda},\btheta,\bxi_d)=\ell\left(\boldsymbol{\Gamma}|\boldsymbol{\lambda},\btheta,\bxi_d\right)+\ell\left( \bZ|\boldsymbol{\lambda}\right).
\end{align*} 
\end{linenomath}
Here $\ell\left(\boldsymbol{\Gamma}|\boldsymbol{\lambda},\btheta,\bxi_d\right)$ and $\ell\left( \bZ|\boldsymbol{\lambda}\right)$ are given by the probability models in \eqref{eqn:LikelihoodLambda} and \eqref{eqn:LikelihoodZ} respectively. With a standard prior specification for the input parameters $\btheta$, the discrepancy parameters in $\bxi_d$, and the latent process $\boldsymbol{\lambda}$, we can define the posterior density for those parameters and carry out Bayesian inference using MCMC.

\subsection{Computational Challenges} \label{section:BinaryChallenges}
The framework described above faces computational and inferential challenges when the number of spatial locations $n$ is large. The emulation step described in Section \ref{section:BinaryEmul} requires estimation of $np$ natural parameters. This is an ill-posed problem since the number of data points in the model output is also $np$. It is not straightforward to find a set of constraints that leads to well-posed parameter inference while still allowing enough flexibility for the resulting process. Even if estimates of the natural parameters are available, emulation still poses significant computational challenges. Emulation involves numerically finding the maximum likelihood estimate of $\bxi_y$, requiring repeated evaluation of a computationally infeasible likelihood function; each likelihood evaluation here involves Cholesky decomposition of the $np\times np$ covariance matrix whose computation scales as $\mathcal{O}(n^3p^3)$. For the example in Section \ref{section:BinaryCalibImplementation}, this calculation translates to a computational cost of $\frac{1}{3} \times p^3 \times n^3=\frac{1}{3} \times 499^3 \times 3,182^3 = 1.33\times10^{18}$ flops, which in turn corresponds to at least hundreds of thousands of hours of computing time for a 3.0 Ghz single core. Moreover, storing $np\times np$ double-precision covariance matrix will require $8n^2p^2=8\times 499^2 \times 3,182^2 = 20,169.33 $ terabytes of memory space. The calibration step is subject to similar challenges. Estimating the input parameters $\btheta$ and the discrepancy parameters $\bxi_d$ requires integrating out the $n$-dimensional discrepancy process $\bdelta$, which makes it difficult to construct a well-mixed MCMC algorithm even for a moderate size of $n$. Moreover, each likelihood evaluation involves Cholesky decomposition of the $n\times n$ covariance matrix and its computational cost scales as $\mathcal{O}(n^3)$. This translates to a computational cost of $\frac{1}{3} \times 3,182^3 =3.59\times 10^{10}$ flops in the example in Section \ref{section:BinaryCalibImplementation}, which corresponds to 30 minutes of computing time for a 3.0 Ghz single core.

\section{Reduced-Dimension Approach for Binary Spatial Data} \label{section:RedDimBinaryCalibration}

We propose a reduced-dimension approach to overcome the computational challenges discussed in the previous section. In the emulation step, we build an emulator for the logistic principal components of the model output to solve the inferential issues in estimating $np$ different natural parameters and the computational issues for dealing with the $np\times np$ covariance matrix. Emulation based on principal components helps us to reduce information loss caused by dimension reduction by focusing on the principal directions for the natural parameters that have the maximum variation across the input parameter settings $\btheta_1,\dots,\btheta_p$. In the calibration step, we use a basis representation approach to model the discrepancy process, which removes the need to integrate out the $n$-dimensional discrepancy process and compute the Cholesky decomposition of the $n\times n$ covariance matrix.

\subsection{Dimension Reduction for Binary Model Output} \label{section:BinaryCalibBinaryprincipal component analysis}
In this subsection, we describe the logistic principal component analysis for binary model output in detail by closely following \citet{lee2010sparse}. For dimension reduction we assume that $\boldsymbol{\Gamma}$ can be written as
\begin{equation} \label{equation:matrix_dim_reduction}
\boldsymbol{\Gamma}=\mathbf{1}_p  \boldsymbol{\mu}^T + \bW \bK_y^T,
\end{equation}
where $\bK_y$ is an $n \times J_y$ orthogonal basis matrix, $\bW$ is the $p \times J_y$ principal component matrix with $(i,j)$th element $w_j(\btheta_i)$, and $\boldsymbol{\mu}$ is the $n \times 1$ mean vector. We rewrite the likelihood in \eqref{equation:BinaryLikelihood} as
\begin{equation} \label{equation:BinaryLikelihoodPC}
L(\bY|\boldsymbol{\mu}, \bK_y,\bW) =\prod_{i=1}^{p} \prod_{j=1}^{n} g(y_{ij}^*(\mu_j+\bw_i \bk_{y,j}^T)),
\end{equation}
where $g(x)=(1+\exp(-x))^{-1}$, $\mu_j$ is the $j$th element of $\boldsymbol{\mu}$, $y_{ij}^*=2y_{ij}-1$, $\bw_i$ is the $i$th row of $\bW$, and $\bk_{y,j}$ is the $j$th row of $\bK_y$.

Following \citet{lee2010sparse}, we estimate the matrix $\bW$ and $\bK_y$ by maximizing the likelihood in \eqref{equation:BinaryLikelihoodPC} using the majorization-minimization (MM) algorithm \citep{lange2000optimization,hunter2004tutorial}. For each iteration of the MM algorithm we minimize the majorizing function instead of the negative likelihood function. The majorizing function is a function that is (i) of the same value as the negative likelihood at the current point of the iteration, (ii) greater than the negative likelihood for all other values of the parameters, and (iii) in an easier form to minimize than the original negative likelihood, such as a quadratic function. As a first step, we note that
\begin{equation}
-\log g(x) \le -\log g(y) - (1- g(y))(x-y) + \frac{1}{8} (x-y)^2.
\end{equation}
for any value of $y$. The equality holds if $x=y$. The proof of this inequality can be found in \citet{de2006principal} (Theorem 4). We can complete the square by adding a term that does not depend on $x$ to rewrite the right hand side as follows;
\begin{equation}
-\log g(x) \le - \log g(y) + \frac{1}{8} \left(x-y-4\left(1-g(y) \right) \right)^2.
\end{equation}
By the fact that $y_{ij}^*$ can take only the value of either +1 or -1, plugging in $y_{ij}^* \gamma_{ij}$ and $y_{ij}^* \gamma_{ij}^{(m)}$ into $x$ and $y$ respectively yields
\begin{linenomath}
\begin{equation}
-\log g(y_{ij}^* \gamma_{ij}) \le - \log g(y_{ij}^* \gamma_{ij}^{(m)}) + \frac{1}{8} \left( \gamma_{ij} - x_{ij}^{(m)} \right)^2,
\end{equation}
\end{linenomath}
where $\gamma_{ij}$ is the $(i,j)$th value of $\boldsymbol{\Gamma}$, $\gamma_{ij}^{(m)}$ is the value of $\gamma_{ij}$ at the $m$th iteration, and $x_{ij}^{(m)}=\gamma_{ij}^{(m)} + 4 y_{ij}^* (1-g(y_{ij}^* \gamma_{ij}^{(m)}))$. Note that \eqref{equation:matrix_dim_reduction} implies that $\gamma_{ij}=\mu_j+\bw_i \bk_{y,j}^T$ and $\gamma_{ij}^{(m)}=\mu_j^{(m)}+\bw_i^{(m)} \bk_{y,j}^{(m)T}$ where $\mu_j^{(m)}$, $\bk_{y,j}^{(m)}$, and $\bw_i^{(m)}$ are the values of $\mu_j$, $\bk_{y,j}$, and $\bw_i$ at the $m$th iteration respectively. Since $- \log g(y_{ij}^* \gamma_{ij}^{(m)}) $ does not depend on $\gamma_{ij}$, the majoring function for the negative log likelihood function in \eqref{equation:BinaryLikelihoodPC} is given by
\begin{linenomath}
\begin{equation*}
\sum_{i=1}^p \sum_{j=1}^n \left( x_{ij}^{(m)} - \mu_j-\bw_i \bk_{y,j}^T \right)^2.
\end{equation*}
\end{linenomath}
For each iteration the minimum of the majorization function occurs at
\begin{linenomath}
\begin{align}
\mu_j =& \frac{1}{p} \sum_{i=1}^p \left( x_{ij}^{(m)}- \bw_i \bk_{y,j}^T \right), \label{equation:MMmean} \\
\bw_i = & (\bK_y^T \bK_y)^{-1} \bK_y^T (\bx_{i.}^{(m)T} - \boldsymbol{\mu}), \label{equation:MMcoefficient}\\
\bk_{y,j} = & (\bW^T \bW)^{-1} \bW^T \left( \bx_{.j}^{(m)}- \mathbf{1}_p \mu_j \right), \label{equation:MMbasis}
\end{align}
\end{linenomath}
where $\bx_{i.}^{(m)}$ and $\bx_{.j}^{(m)}$ are the $i$th row and the $j$th column of $\bX^{(m)}$ respectively and $\bX^{(m)}$ is a $p \times n$ matrix having $x_{ij}^{(m)}$ as its $(i,j)$th element. Cycling through \eqref{equation:MMmean} - \eqref{equation:MMbasis} results in a local maxima of the likelihood function in \eqref{equation:BinaryLikelihoodPC}. Results related to convergence for this algorithm can be found in \citet{hunter2004tutorial}.  Of course, as in other optimization algorithms one needs to be careful about possible stationary points. We summarize the algorithm as follows.
\begin{enumerate}
\item Initialization: Set $m=1$ and choose starting points $\bmu^{(1)}=(\mu_1^{(1)},\dots,\mu_n^{(1)})^T$, $\bW^{(1)}=\left(\bw_1^{(1)T},\dots,\bw_p^{(1)T} \right)^T$ and $\bK_y^{(1)}=\left(\bk_{y,1}^{(1)T},\dots,\bk_{y,n}^{(1)T} \right)^T$.
\item Compute $\gamma_{ij}^{(m)}=\mu_j^{(m)}+\bw_i^{(m)} \bk_{y,j}^{(m)T}$, $x_{ij}^{(m)}=\gamma_{ij}^{(m)} + 4 y_{ij}^* (1-\pi(y_{ij}^* \gamma_{ij}^{(m)}))$ and let $\bX^{(m)}=\left\{ x_{ij}^{(m)} \right\} $.
\item Update $\boldsymbol{\mu}^{(m+1)}= \frac{1}{p} \left( \bX^{(m)} - \bW^{(m)} \bK_y^{(m)T} \right)^T \mathbf{1}_p$.
\item Compute $\bW^{*}=(\bX^{(m)}- \mathbf{1}_p  \boldsymbol{\mu}^{(m+1)T})^T\bK_y^{(m)} \left( \bK_y^{(m)T} \bK_y^{(m)} \right)^{-1}$ and find its singular value decomposition 
$\bW^*=\bU \bLambda \bV^T$. Update $\bW^{(m+1)}=\bU \bV^T$.
\item Update $\bK_y^{(m+1)}=\left( \bX^{(m)}- \mathbf{1}_p \boldsymbol{\mu}^{(m+1)T}\right)^T \bW^{(m+1)}$.
\item Set $m+1$ and repeat from 2 until convergence.
\end{enumerate}

\subsection{Emulation Using Logistic PCs} \label{subsection:EmulationLogisticPC}
We fit a Gaussian process for each column of $\bW$ separately to construct a $J_y$-dimensional emulator process. We denote the value of the $j$th principal component as $w_j(\btheta)$ and set up a Gaussian process for each $j$th column with the following exponential covariance function for any parameter settings $\btheta$ and $\btheta'$:
\begin{equation} \label{eqn:CovFunc}
\Cov \left(w_{j}(\btheta),w_{j}(\btheta') \right) = \kappa_{j}\exp\left(-\sum_{i=1}^{d} \frac{|\theta_{i}-\theta_{i}'|}{\phi_{ij}} \right)+\zeta_{j} 1(\btheta=\btheta'),
\end{equation}
with the partial sill $ \kappa_{j}$, the range parameters $\phi_{ij}$, the nugget parameter $\zeta_{j}$, and the $i$th element of $\btheta$, $\theta_{i}$. For each $j$th principal component, we estimate the parameters $\kappa_{j}$, $\phi_{1j},\dots,\phi_{dj}$, and $\zeta_{j}$ by maximizing the log likelihood function $\ell(w_{j}(\btheta_1),\dots,w_{j}(\btheta_p)|\kappa_{j},\phi_{1j},\dots,\phi_{dj}, \zeta_{j})$ given by the Gaussian process for $w_{j}(\btheta_1),\dots,w_{j}(\btheta_p)$ defined by the covariance function in \eqref{eqn:CovFunc}. Here we use the exponential covariance function because we expect that there is not enough information in the binary data to infer the smoothness parameter in Mat\'{e}rn. Our choice of smoothness parameter 0.5 for the Matern covariance function (therefore an exponential covariance function) is a conservative one. We have also conducted a leave-10-percent-out cross validation experiment and found that an emulator with the exponential covariance function shows a better interpolation performance than the squared exponential covariance function. The computational cost for each likelihood evaluation is $\frac{1}{3}p^3$ flops, which is $\frac{1}{3} \times 499^3\approx4.14 \times 10^7$ flops in our application.

The $J_y$-dimensional emulator process $\boldeta\left(\btheta,\bW\right)$ for the logistic PC scores at any new parameter value $\btheta$ is obtained as usual from the conditional distribution. Then the $n$-dimensional emulator process for the natural parameters at $\btheta$ is
\begin{equation} \label{equation:GammaTheta}
\boldsymbol{\gamma}(\btheta)=\boldsymbol{\mu}+\bK_y \boldeta\left(\btheta,\bW\right),
\end{equation}
and the corresponding probability for the predicted model output at each location to be one is
\begin{linenomath}
\begin{equation*}
p_j(\btheta)=P(Y(\btheta,\bs_j)=1)=\frac{\exp\left(\gamma_j(\btheta)\right)}{1+\exp\left(\gamma_j(\btheta)\right)},
\end{equation*}
\end{linenomath}
where $\gamma_j(\btheta)$ is the $j$th element of $\boldsymbol{\gamma}(\btheta)$ in \eqref{equation:GammaTheta}.

\subsection{Calibration Using Logistic PCs} \label{section:BinaryCalibCalibration}
Once the emulator for the logistic PCs is available, we set up the following calibration model for the $n$-dimensional vector of the natural parameters $\boldsymbol{\lambda}$ for observational data using the emulator $\bmu+\bK_y \boldeta\left(\btheta,\bW\right)$:
\begin{equation} \label{eq:lambda_model}
\boldsymbol{\lambda} =\boldsymbol{\mu}+ \bK_y \boldeta+ \bdelta,
\end{equation}
where $\btheta$ is the best fit value for the natural parameters for observational data, $\boldeta=\boldeta\left(\btheta,\bW\right)$ is the emulated natural parameter vectors at the value of input parameter $\btheta$, $\bdelta$ is the discrepancy term that represents structural error between the natural parameters for the model output and those for the observational data. The intercept $\bmu$ and the basis matrix $\bK_y$ are given from the emulation stage. To get around the challenges in integrating out $\bdelta$ described above, we use a basis representation for the discrepancy term such that
\begin{linenomath}
\begin{equation*}
\bdelta=\bK_d \mathbf{v},
\end{equation*}
\end{linenomath}
with the $n\times J_d$ basis matrix $\bK_d$ and the $J_d$-dimensional random coefficient vector $\mathbf{v} \sim N(\mathbf{0},\sigma^2_d \mathbf{I}_{J_d})$. Similar to the independence assumption between the emulated vector and the discrepancy term made for the model in \eqref{eqn:LikelihoodLambda}, we assume that $(\mathbf{v},\sigma^2_d)$ is  independent of $(\boldeta, \btheta, \bW)$.

 The posterior density given the observational data $\bZ$ (see Supplementary Section A1 for details about its derivation) can be written as
\begin{linenomath}
\begin{align} \label{eq:posterior}
\pi(\boldeta,\btheta,\bv,\sigma_d|\bZ,\bW)\propto L\left( \bZ|\boldeta,\bv\right) f\left(\boldeta|\btheta,\bW\right) f\left(\bv|\sigma_d\right)f(\btheta) f(\sigma_d),
\end{align}  
\end{linenomath}
where the likelihood function $L\left( \bZ|\boldeta,\bv\right)$ is 
\begin{linenomath}
\begin{align*}
L\left( \bZ|\boldeta,\bv\right) &\propto \prod_{j=1}^n \left( \frac{\exp(\lambda_j)}{1+\exp(\lambda_j)} \right)^{Z(\bs_j)} \left( \frac{1}{1+\exp(\lambda_j)} \right)^{1-Z(\bs_j)},
\end{align*}
\end{linenomath}
with $\lambda_j$ being the $j$th element of $\boldsymbol{\lambda}=\boldsymbol{\mu}+ \bK_y \boldeta+ \bdelta$, and $ f\left(\boldeta|\btheta, \bW \right)$ and  $f\left(\bv|\sigma_d\right)$ are 
\begin{linenomath}
\begin{align*}
 f\left(\boldeta|\btheta, \bW\right) &\propto \prod_{k=1}^{J_y} \frac{1}{\sigma_{\eta,k} (\btheta)} \exp\left(-\frac{(\eta_{k} - \mu_{\eta,k}(\btheta))^2}{2\sigma_{\eta,k}(\btheta)^2}   \right), \\
f\left(\bv|\sigma_d\right) &\propto  \prod_{l=1}^{J_d} \frac{1}{\sigma_d} \exp\left(-\frac{v_l^2}{2\sigma_d^2} \right),
\end{align*}
\end{linenomath}
where $\mu_{\eta,k}(\btheta)$ and $\sigma_{\eta,k}^2(\btheta)$ are the mean and the variance of the $k$th element of $\boldeta(\btheta,\bW)$, $v_l$ is the $l$th element of $\bv$, and  $\sigma_d^2$ is the common variance for $v_1\dots v_{J_d}$. For the priors we assume that $f(\btheta)$ is a $d$-dimensional uniform density whose support covers all plausible values of $\btheta$ and $f(\sigma_d)$ is an inverse-Gamma density with the shape parameter $a_d$ and the scale parameter $b_d$. The functions $L\left( \bZ|\boldeta,\bv\right)$, $f\left(\boldeta|\btheta,\bW\right)$, and $f\left(\bv|\sigma_d\right)$ are based on proper densities, hence  the posterior is proper as long as the priors $f(\btheta)$ and $f(\sigma_d)$ are proper. The observations $Z(\bs_1),\dots,Z(\bs_n)$ are conditionally independent given the values of the random coefficients $\boldeta$ and $\bv$ and the dependence between them is accounted for through the basis matrices $\bK_y$ and $\bK_d$. Based on the specifed posterior one can infer the parameters via MCMC.

Finding the basis $\bK_d$ that can precisely represent the discrepancy process with a small number of basis vectors is important for computational tractability. To this end, we use the following procedure that is inspired by the definition of the discrepancy term:\begin{enumerate}
\item[1.] For each location $\bs_j$, compute the signed proportion of mismatch between the model output and the observational data as $r_j=\sum_{i=1}^p  \mbox{sgn}(Y(\btheta_i,\bs_j)-Z(\bs_j)) I(Y(\btheta_i,\bs_j)\ne Z(\bs_j)) /p$ where $\mbox{sgn}$ is the sign function. 
\item[2.] Set $J_d=1$ and hence $\bK_d$ is reduced to a $n$-dimensional vector $\mathbf{k}_d$. Define the $j$th element of $\bk_d$ as $\log(\frac{1+r_j}{1-r_j})$ if $|r_j|>c$, or as $0$ if $|r_j|<c$. The random coefficient vector $\bv$ is also reduced to a univariate random variable $v\sim N(0,\sigma_d^2)$.
\end{enumerate}
The goal of step 1 is to capture a common discrepancy pattern that is persistent across all parameter settings. For this procedure to work the design points $\btheta_1,\dots,\btheta_p$ have to be a representative sample of plausible values of $\btheta$, and we expect that this holds for the Latin hypercube design that we are using here. In step 2 we apply logit transformation $\log \left( \frac{1+r_j}{1-r_j} \right)$ for $-1\le r_j \le 1$ to translate the identified discrepancy pattern into logit scale since the discrepancy term $\bk_d v$ needs to be specified for the natural parameters. Note that the cut-off value $c$ in step 2 needs to be determined a priori. Choosing a value of $c$ that is too large results in loss of important discrepancy patterns in $\bk_d$ and, on the other hand, a value of $c$ that is too small causes identifiability issues in estimating $\btheta$.  In the simulated examples described in Section \ref{subsection:SimulatedExamples}, we have found that the value of $c=0.5$ strikes a nice balance, and yields a discrepancy basis that capture the true simulated discrepancy patterns reasonably well (See Figure A1 in the supplement). This value of $c$ is a heuristic but it appears to work well for the variety of examples to which we have applied our methods. We set $c=0.5$ for the remainder of this paper. Note also that $\log(\frac{1+r_j}{1-r_j})$ can take a positive or negative infinite when $r_j=1$ or $r_j=-1$. To prevent this, we set $r_j=0.999\times \mbox{sgn}(r_j)$, where $\mbox{sgn}(r_j)$ is the sign of $r_j$, when computing $\log(\frac{1+r_j}{1-r_j})$ if $|r_j|$ is greater than 0.999.

\subsection{Procedure Summary}

Let $\bY$ be an $n\times p$ matrix of the computer model output where the columns represent different spatial locations and the rows represent different parameter settings and $\bZ$ denotes an $n$-dimensional vector of observational data (defined in Section \ref{section:BinaryCalibEmul}). We summarize the overall emulation-calibration procedure described in this section as follows:
\begin{enumerate}
\item Dimension-reduction of computer model output: Compute logistic principal components of $\bY$ to reduce the dimensionality from $n$ to $J_y$ (details in Section \ref{section:BinaryCalibBinaryprincipal component analysis}) by maximizing the likelihood function in \eqref{equation:BinaryLikelihoodPC}. We store the principal component scores in a $p \times J_y$ matrix $\bW$, the principal component basis in an $n \times J_y$ matrix $\bK_y$, and the location means in an $n$-dimensional vector $\bmu$ (see \eqref{equation:matrix_dim_reduction} for their definition).
\item Gaussian process-based univariate emulators: We construct a one-dimensional Gaussian process emulator for each column of $\bW$ (details in Section \ref{subsection:EmulationLogisticPC}) with the covariance function in \eqref{eqn:CovFunc}. Denote the $J_y$-dimensional vector of the emulated principal components at a new parameter value $\btheta$ by $\boldeta(\btheta,\bW)$. 
\item Constructing a discrepancy basis: Find the discrepancy basis vector $\bk_d$ by comparing each row of $\bY$ and $\bZ$ (details in Section \ref{section:BinaryCalibCalibration}). Denote the coefficient for the discrepancy basis vector by $v$.
\item  Calibration using Markov chain Monte Carlo: Based on the posterior density in \eqref{eq:posterior} infer the input parameters $\btheta$, the principal components $\boldeta$, the discrepancy coefficient $v$, and the discrepancy variance $\sigma_d^2$ using the Metropolis-Hastings algorithm (details in Section \ref{section:BinaryCalibCalibration}).
\end{enumerate} 
Our procedure requires estimating only $J_y+d+2$ parameters in the calibration step. For our ice sheet model calibration problem, this corresponds to only $10+10+2=22$ parameters. Note that it would require estimating $n+d+1=3,182+10+1=3,193$ parameters without dimension reduction.

\section{Implementation Details and Results} \label{section:BinaryCalibImplementation}

In this section we study the application of our approach to simulated examples and then calibrate the PSU3D-ICE model based on the real observational data. We provide descriptions for each of the calibrated parameters as well as scientific interpretations of our results.

\subsection{Simulated Examples} \label{subsection:SimulatedExamples}

We now provide a discussion of our investigation of our approach in the context of simulated examples. We describe our construction of simulated examples and  results from the applications of our methods to these examples, along with implementation details for our method. We apply our method to (i) a simple synthetic ice sheet model example to study the properties of our method, and to (ii) the PSU3D-ICE to estimate the input parameters of the model and make projections for future ice sheet volume changes in the Amundsen basin region.

\bigskip
\textit{Simple Numerical Example.}  \label{subsection:BinarySimulatedExample} We begin with a simple numerical example. Unlike the real ice sheet model calibration problem, this example does not have complicating issues such as lack of identifiability of input parameters or sparsity of design points in the parameter space, and hence it provides a nice first test for our calibration methods.

We defined the model output as $Y(\btheta,\bs)=I\left(U(\btheta,\bs)\right)$ where
\begin{linenomath}
\begin{equation*}
U(\btheta,\bs)=\sqrt{1-s_1^2-\left(\frac{s_2}{1.5}\right)^2-\theta_1+\theta_2(s_2+1.5)},
\end{equation*}
\end{linenomath}
with $\bs=(s_1,s_2)$ and $\btheta=(\theta_1,\theta_2)$ for $0.3<\theta_1<0.65$, $0<\theta_2<0.2$, $0<s_1<1$, and $-1.5<s_2<1.5$. We form an ensemble of model outputs with 100 runs where the design points for $(\theta_1,\theta_2)$ are on a regular $10 \times 10$ lattice covering the entire parameter space $\boldsymbol{\Theta}=[0.3,0.65]\times[0,0.2]$. Each model run $\bY(\btheta_i)$ is a spatial pattern on a $30\times30$ regular lattice that covers the entire spatial field $\mathcal{S}=[-1,1]\times[-1.5,1.5]$ and hence $n=900$. See Figure A2 in the Supplement for 16 sample model runs. The first parameter $\theta_1$ controls the overall size of the grey area and the second parameter $\theta_2$ changes both the shape and the size of the grey area. We set the number of principal components to $J_y=10$ based on leave-10-percent-out cross-validation; using more than 10 principal components does not significantly improve the emulation performance in our cross-validation experiments. The overall misclassification rate is less than 5\%. We construct a synthetic observational data set by choosing the model run with $(\theta_1,\theta_2)=(0.494,0.089)$ as the synthetic truth and contaminating the corresponding model output with noise generated from the discrepancy process, defined as a zero-mean Gaussian process with an exponential covariance function with the partial sill of $1.5$, the range of $0.03$ and the nugget of $0.00001$. This affects the resulting binary pattern considerably by changing roughly 10\% of the original output. We repeat the same experiment with 10 different realizations from the discrepancy process and the results are summarized in Figure A3 in the Supplement. The results show that, on average, we can recover the true input parameter setting reasonably well in this simple numerical setting.

\bigskip
\textit{Realistic Example Using the Ice Sheet Model PSU3D-ICE.} \label{section:CalibrationIce}
We now work with a more realistic synthetic observational data set to verify that our calibration approach works well in the context of the real ice sheet model calibration problem. Here we use the model runs from the PSU3D-ICE model described in Section \ref{section:DataDescription} to generate the example. This estimation problem is more challenging than the example described in the previous section because it involves complicated interactions between the input parameters. Also, the input parameter settings are much more sparsely distributed in the parameter space (499 ensemble members in 10 dimensional space), which potentially makes emulation, as well as calibration, more difficult.

We select model run \#394 as the synthetic truth, because it is among the model runs closest to the observational data. We count the number of pixels where the observational data and the model output have the same value, i.e. we compute  $\sum_{j=1}^n I(Z(\bs_j)=Y(\btheta_i,\bs_j))$ for each $\btheta_i$ where $I(.)$ is the indicator function. A model run is close to the observations if this count is large. Since in reality the model-observation discrepancy operates in terms of the actual ice thickness patterns (the binary ice-no ice patterns are derived from these data), we construct the discrepancy pattern in terms of ice thickness and translate it into a binary spatial pattern. We use the following steps to construct a realistic model-observation discrepancy. 

\begin{enumerate}
\item By comparing the original model runs and the observational data that are spatial patterns of ice thickness, find the top 60\% model runs with the smallest root mean squared error (RMSE) values.
\item For each spatial location, take the average of the difference between the observational data and the model runs selected in step 1 to find the common discrepancy pattern in ice thickness.
\item Subtract common discrepancy pattern found in step 2 from the ice thickness output corresponding to the synthetic truth. The resulting pattern is synthetic observational data for ice thickness.
\item Dichotomize the ice thickness pattern computed in step 3 at 0 to obtain synthetic observational data for ice-no ice pattern.
\end{enumerate}
The procedure described in Section \ref{section:BinaryCalibCalibration} recovers the superimposed discrepancy reasonably well (Figure A1 in the Supplement, lower panels).

\begin{figure}
\centering
\includegraphics[scale=0.41]{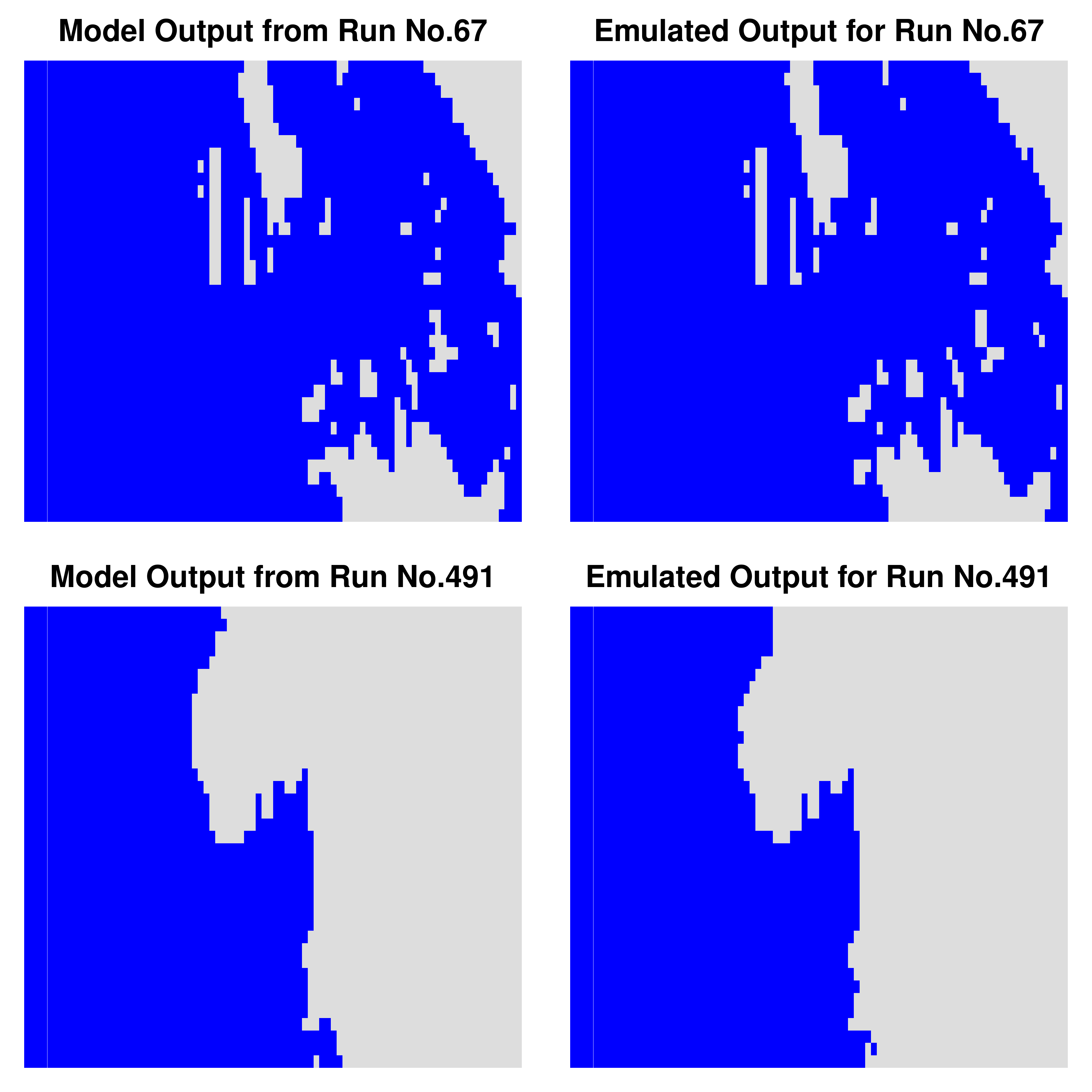}
\caption{Two examples of leave-10-percent-out cross-validation results to study the performance of our logistic PCA based emulator. The gray pixels show ice-covered locations. (a) Comparison between the original (left) and the emulated (right) ice coverage patterns for model run no. 67. (b) The same comparison for model run no. 491. The graphical comparison shows that the emulated model output (right panel) approximates the original model output (left panel) fairly closely. Similar results hold for the other model runs.}
\label{fig:BinaryCrossValidation}
\end{figure}

We choose to use 10 principal components for dimension reduction based on the same cross-validation criteria described in the previous section. The leave-10-percent-out cross-validation confirms that our emulator can predict the model output precisely (Figure \ref{fig:BinaryCrossValidation}), with an overall error rate of 4\%.  To translate the probabilities $p_1(\btheta),\dots,p_n(\btheta)$ from our emulator into predictions for the binary outcome, we simply dichotomize them using 0.5 as a threshold. We give a non-informative prior $IG(2,3)$ to $\sigma_d^2$ and flat priors to the input parameters in $\btheta$. The range of each flat prior is restricted to the range for the design points $\btheta_1,\dots,\btheta_p$. 
Using the standard Metropolis-Hastings algorithm we generate MCMC sample for $\btheta$ while integrating out $\boldeta$, $v$, and $\sigma_d^2$. For each experiment we obtain MCMC chain with 1,000,000 iterations. Looking at MCMC standard errors \citep{jones2006fixed,flegal2008markov} suggests that the sample size was large enough to estimate posterior means while accounting for autocorrelations. We also overlaid the marginal density plots obtained after half the run over the marginal density plots obtained after the entire run, confirming that the plots do not show significant changes and the results are therefore stable. The computation takes about 120 hours for the entire MCMC run on a system with Intel Xeon E5450 Quad-Core 3.0 GHz..

\begin{figure}
\centering
\includegraphics[scale=0.6]{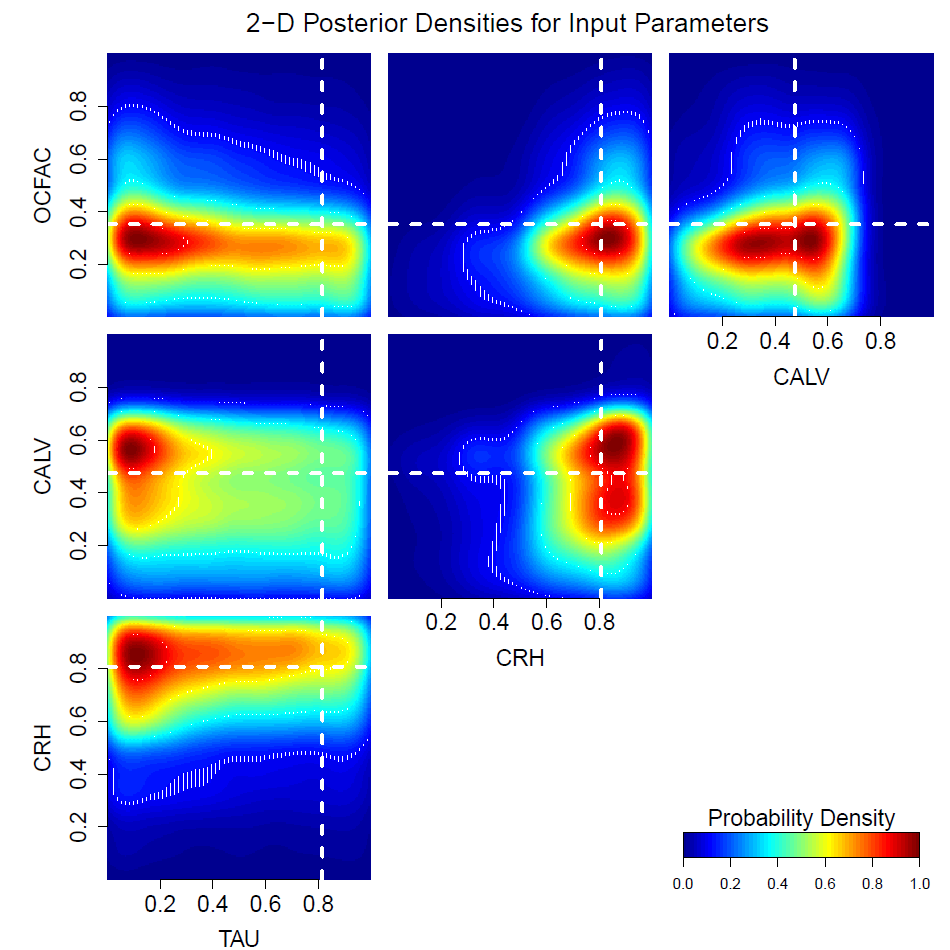}
\caption{The pairwise joint posterior densities for the input parameters from the calibration results for the simulated example in Section \ref{section:CalibrationIce}. We plot only the following four parameters that we are most interested in: ocean sub-ice-shelf melt factor (OCFAC), calving factor (CALV), basal sliding coefficient (CRH), and asthenospheric relaxation \textit{e}-folding time (TAU) (see Section \ref{subsection:ObsCalibration} for parameter descriptions). The white dashed lines show the input parameter settings for the synthetic truth. The marginal densities  peak around the assumed true values for the parameters. The lone exception is TAU, indicating that the modern ice coverage pattern is not informative about this parameter.}
\label{fig:BinaryIceResults}
\end{figure}

The pairwise joint densities of input parameters are shown in Figure \ref{fig:BinaryIceResults}. Since the scientific interest is mainly in the four parameters, ocean sub-ice-shelf melt factor (OCFAC), calving factor (CALV), basal sliding coefficient (CRH), and asthenospheric relaxation \textit{e}-folding time (TAU), we display the results for only these parameters (see Section \ref{subsection:ObsCalibration} below for the reason). The density plots indicate that the marginal posterior densities for the first three densities peak around the true parameter values, but the marginal densities for TAU do not recover the true values and show large dispersion comparing to the other parameters. We hypothesize that the binary spatial patterns are not informative about TAU. The probable reason for parameter TAU not being well constrained is that the statistical analysis above is limited to calibration with the modern observed ice state. TAU is the \textit{e}-folding time scale of asthenospheric relaxation below the varying ice load (see Table \ref{table:parameters} and Section \ref{subsection:ObsCalibration} below), and its primary effect on ice evolution occurs during the main deglacial retreat phase, 12 to 8 ka. The ice retreat is nearly complete by 8 ka in most of our model runs, leaving ample time for the asthenosphere to relax towards the modern state, regardless of the exact value of TAU. Calibration and statistical techniques using past grounding-line positions vs. those estimated from geologic data over the last 15 kyr \citep{larter2014reconstruction} will likely be able to diagnose and constrain this important parameter.

Using the MCMC chain for the input parameters, we generate projections for WAIS volume change. Here we present projection results for 500 and 1000 years from the present day. We build an emulator for ice volume changes generated from the PSU3D-ICE and convert the MCMC chain for input parameters into projections using the emulator. The results in Figure \ref{fig:IceVolumeProjection} show the predictive distribution of the ice volume change projections corresponding to the estimated posterior density. For comparison, we have also computed the predictive density by assuming that all the parameter settings in the model ensemble are equally likely. We compare these densities with the assumed true ice volume change projection values produced by the parameter settings for the synthetic truth. For both time points (500 and 1000 years from the present) the 95\% prediction intervals cover the synthetic true ice volume change projections. The mode of the predictive density matches well with the true ice volume change projection for 500 years, but the mode for the ice volume change for 1000 years does not match with the assumed true value. We hypothesize that this is due to the large uncertainty for TAU, which controls the long term ice sheet behavior. We provide a longer discussion of this in the next section.

\subsection{Calibration of PSU3D-ICE Using Observational Data} \label{subsection:ObsCalibration} 

\begin{figure}
\centering
\includegraphics[scale=0.6]{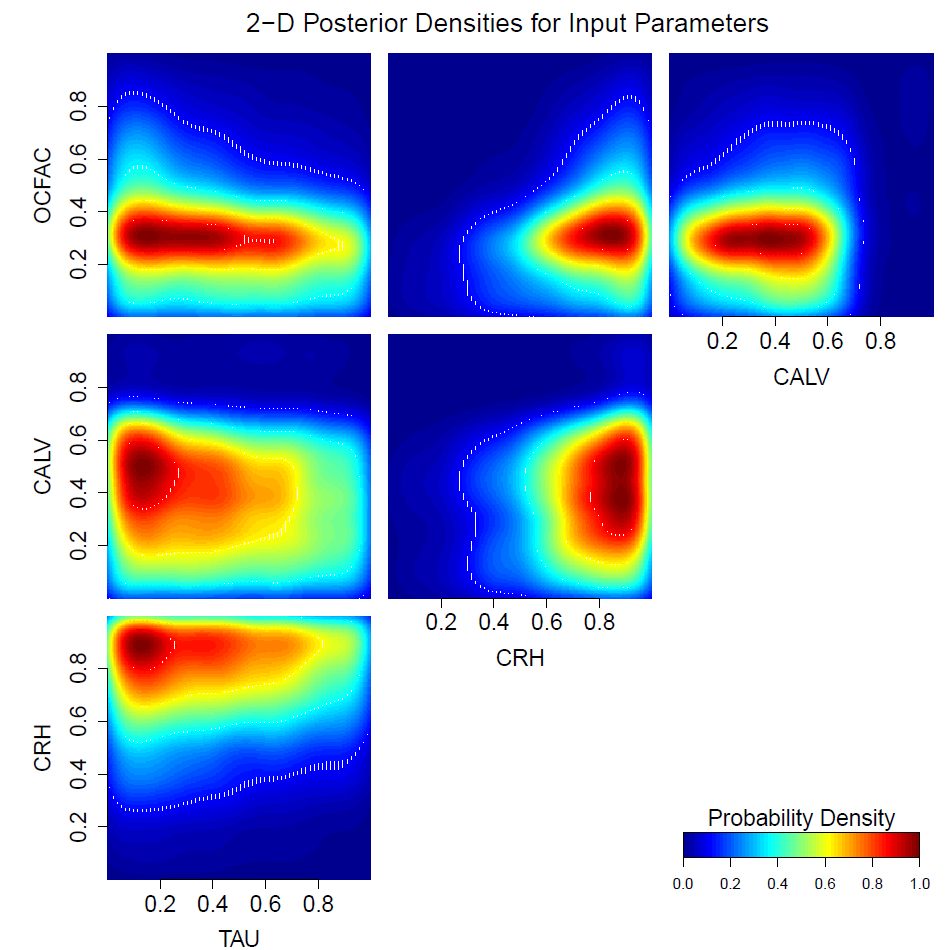}
\caption{The pairwise joint posterior densities for the input parameters from the calibration results based on the real observational data described in Section \ref{section:DataDescription}. Again we plot only the following four important parameters: ocean sub-ice-shelf melt factor (OCFAC), calving factor (CALV), basal sliding coefficient (CRH), and asthenospheric relaxation \textit{e}-folding time (TAU) (see Section \ref{subsection:ObsCalibration} for parameter descriptions). Similar to the results in Figure \ref{fig:BinaryIceResults}, the posterior densities have tightly constrained high density regions except for TAU.}
\label{fig:BinaryIceResultsObs}
\end{figure}

\begin{figure}
\centering
\includegraphics[scale=0.4]{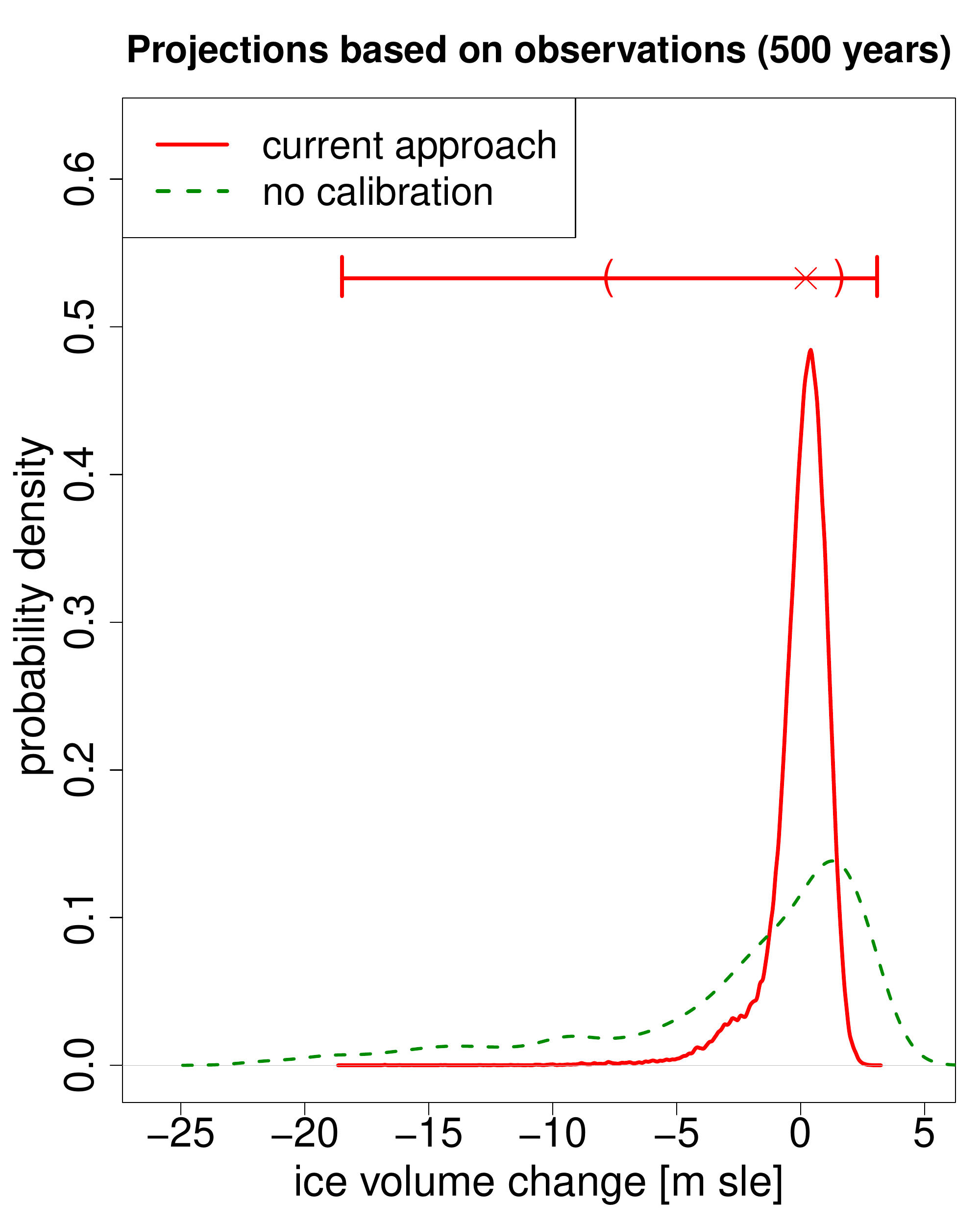}
\includegraphics[scale=0.4]{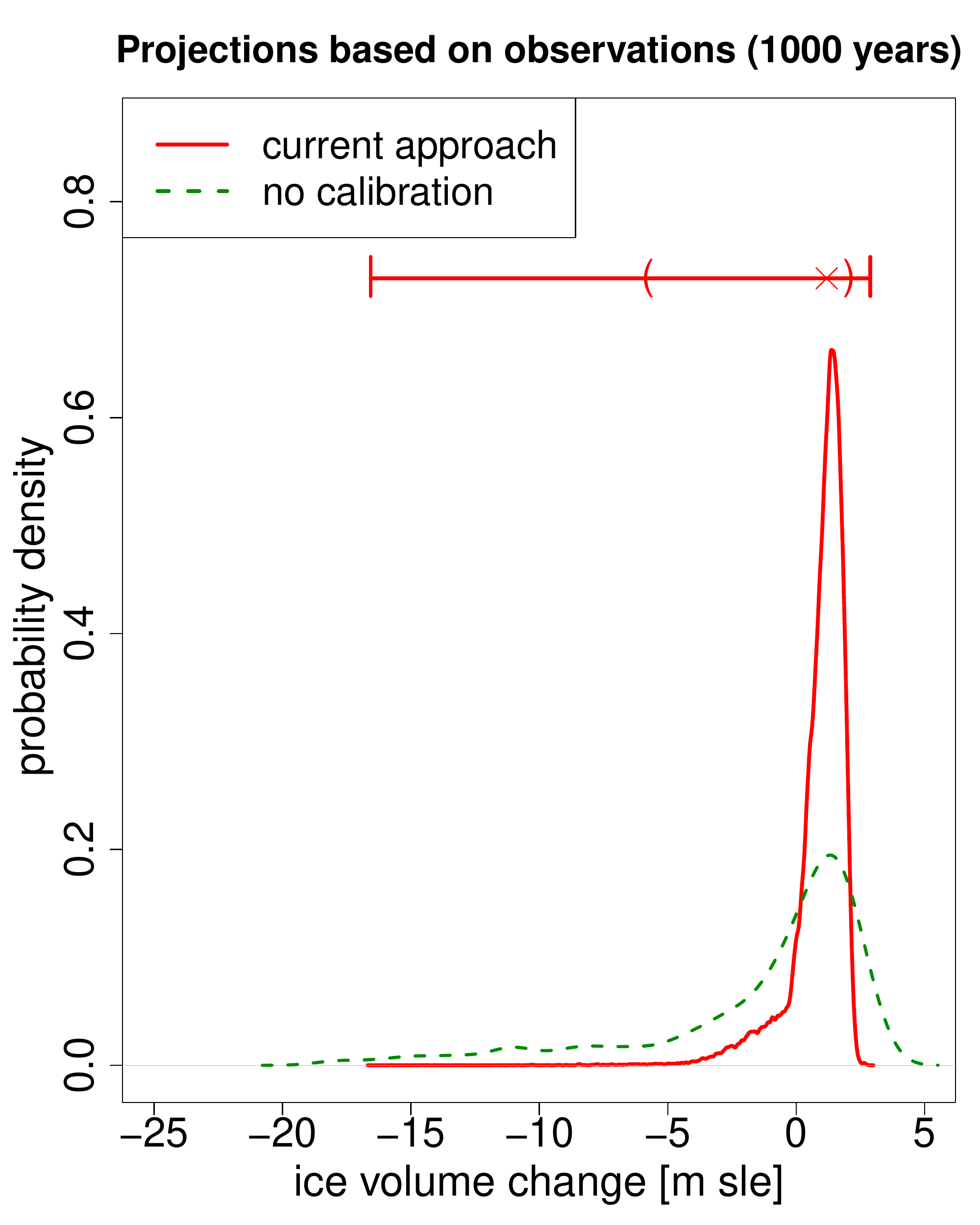}
\caption{The ice volume change projections for 500 years (left panel) and 1000 years (right panel) based on the calibration results (solid red lines) using the real observational data shown in Figure \ref{fig:BinaryIceResultsObs} and no calibration (dashed green lines). The parentheses on the horizontal bars represent 95\% prediction intervals and the vertical bars represent the ranges.}
\label{fig:IceVolumeProjectionObs}
\end{figure}

We now calibrate PSU3D-ICE using the observational data. The calibration results based on the observations from the Bedmap2 dataset \citep{fretwell2013bedmap2} are shown in Figure \ref{fig:BinaryIceResultsObs} and \ref{fig:IceVolumeProjectionObs}. 

\bigskip
\textit{Parameter Description.} 
 Out of the 10 ice-sheet model parameters that were varied and inferred we focus on four important parameters. These parameters are highly uncertain even though they have strong effects on model behavior. Most parameters and inputs that concern interior grounded-ice processes, inland of the modern grounding zone, are relatively well known, based on modern data and well established in previous model-data studies. Examples are ice rheology, surface accumulation, and basal sliding through inverse procedures. However, processes that affect the floating ice shelves, and also properties of past grounded ice that advanced out over the Antarctic continental shelf at glacial maxima, are much less well
known, and we focus on important parameters in this category. The 4 parameters and processes are:
\begin{itemize}
\item Ocean sub-ice-shelf melt factor (OCFAC): Oceanic melting at the base of floating ice shelves, due to warm waters flowing from the open ocean into the cavities below the ice shelves. This affects ice shelf thickness, and changes buttressing of interior ice. Recent increases in ocean melt rates are considered to be the main cause of drastic ongoing ice retreat in the ASE sector of WAIS \citep{pritchard2012antarctic,rignot2002rapid}. 
For small (large) OCFAC values, oceanic melting is reduced (increased), ice shelves thicken (thin), discharge of interior ice across the grounding line decreases (increases), and grounding
lines tend to advance (retreat).

\item Calving factor (CALV):  Calving of icebergs at the oceanic edge of floating shelves. This process is not well understood, probably depending on pre-existing fracture regimes, large-scale stresses, and hydrologic conditions, yet has important effects on ice-shelf extent and feedback on buttressing and internal ice, both for the modern state and also for past and future variations. There is little consensus on calving parameterizations. We use a common approach based on (highly parameterized) crevasse depths and their ratio to ice thickness (Benn et al. 2007; Nick et al., 2010). 
For small (large) CALV, calving is decreased (increased), producing more (less) extensive floating shelves, and greater (lesser) buttressing of interior ice.

\item Basal sliding coefficient (CRH): Basal sliding of grounded ice at the interface with bedrock. For beds at the ice pressure-melt point, for a given driving stress, sliding depends mainly on whether the bed is hard and non-deformable (crystalline rock, sticky, slow sliding) or soft and deformable (unconsolidated sediment or till, slippery, fast sliding). The coefficient values in the basal sliding law relating basal stress to velocity have strong effects on the overall ice
profile shape and thicknesses, and affect the overall sensitivity to climate change.  Coefficients under modern grounded ice are deduced by inverse methods \citep{pollard2012description,morlighem2013inversion}, but they are relatively unconstrained for modern oceanic beds, where grounded ice advanced during the last glacial maximum 20 to 15 ka.  Most oceanic areas around Antarctica are covered in deformable sediment today, due to marine sedimentation and
to generation and transport by previous ice advances. For these regions, there is a wide plausible range of relatively large sliding coefficients that have significant effects on model results. Here, we vary the sliding coefficient CRH just for modern oceanic areas.

\item Asthenospheric relaxation \textit{e}-folding time (TAU): The ice sheet evolution on $10^3$ year timescales depends strongly on the bedrock response to varying ice load. During deglacial retreat, the bedrock rebounds upwards due to the retreating ice load, but with an $O(10^3)$ year lag, and this lag has an important positive feedback via deeper bedrock and grounding-line depths. As in many ice models, this bedrock response is represented by a simple Earth model consisting of an elastic plate (lithosphere) over a local \textit{e}-folding lagged 
relaxation towards isostatic equilibrium (asthenosphere). Based on 
more sophisticated Earth models, the asthenospheric \textit{e}-folding time scale is commonly set to 3 kyr \citep[e.g.,][]{gomez2013three}, but note that recent studies suggest considerably shorter time scales for some West Antarctic regions \citep{whitehouse2012new}. 

\end{itemize}

\begin{figure}
\centering
\includegraphics[scale=0.70]{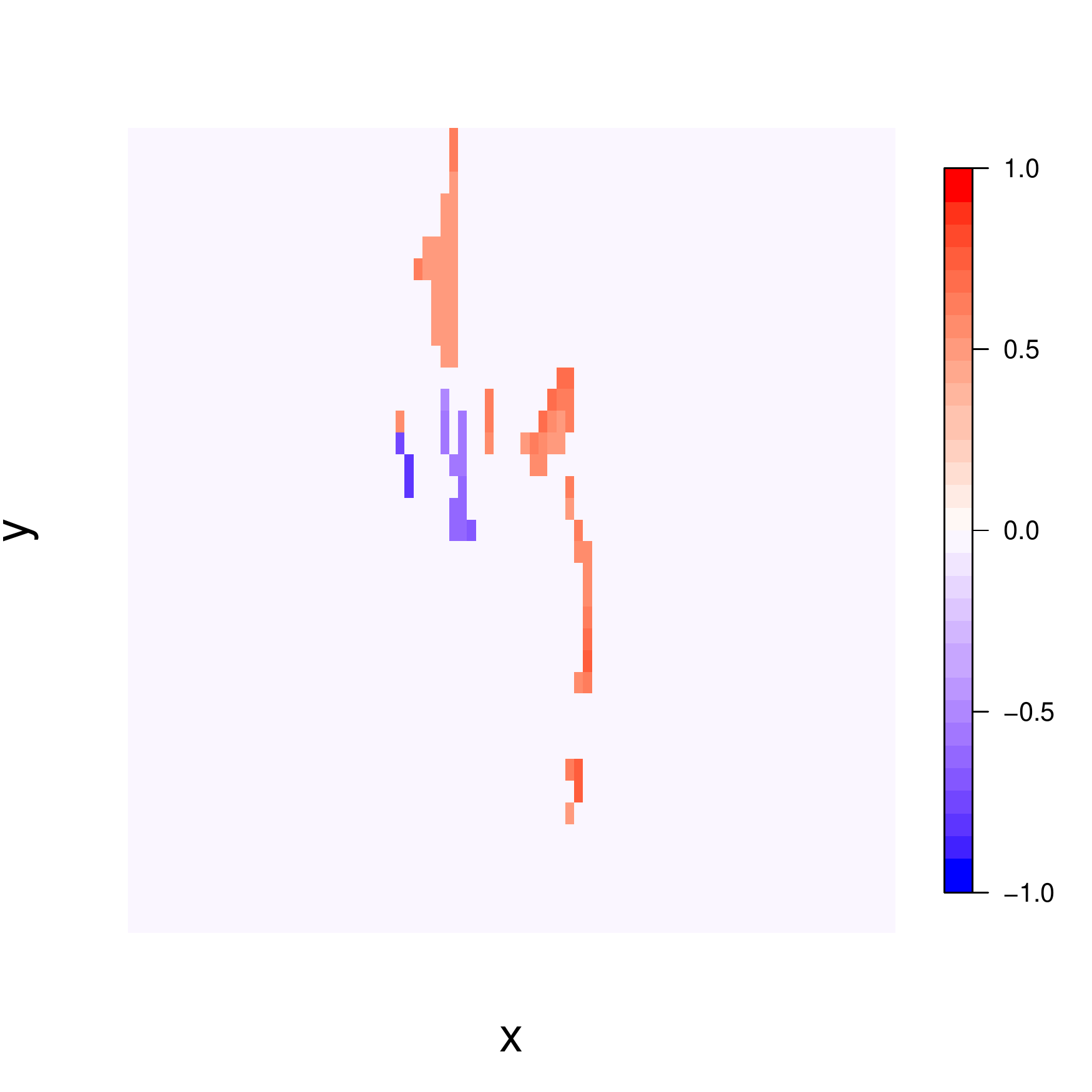}
\caption{Estimated discrepancy between PSU3D-ICE model and the Bedmap2 observational dataset \citep{fretwell2013bedmap2} by the procedure described in Section \ref{section:BinaryCalibCalibration}. }
\label{fig:EstimatedDisc}
\end{figure}

\bigskip
\textit{Results.} 
The estimated discrepancy pattern by the procedure described in Section \ref{section:BinaryCalibCalibration} is shown in Figure \ref{fig:EstimatedDisc}. Much of the discrepancy pattern in Figure \ref{fig:EstimatedDisc} is due to a known shortcoming of the ice model that tends to fill narrow bays and passages between mainland and nearby islands with grounded ice, where there is only floating ice in the real world. This is partly due to the coarse model grid size (20 km) not sufficiently resolving the passages, and also presumably due to the lack of fine-scale variations in oceanic basal melting that could well exist in these passages due to local ocean currents.

As seen in the pairwise density plots in Figure \ref{fig:BinaryIceResultsObs}, the best values of ocean melt coefficient (OCFAC) and calving factor (CALV) are roughly in their mid-range, i.e., quite close to the values in the nominal model that relied on prior tuning by the model developers \citep{pollard2012description}. There is a slight indication that OCFAC should be somewhat smaller than the value found by \citet{pollard2012description}. The best CALV value, 0.5, corresponds to an unchanged nominal value.

The bedrock (asthenospheric) \textit{e}-folding rebound time TAU is not constrained well in Figure \ref{fig:BinaryIceResultsObs}. Although the uppermost values (corresponding to 6 to 7 kyr in the original scale) are excluded, values below 6 kyr have similar probabilities. As discussed above, this is expected because TAU primarily affects past ice distribution during deglaciation, with little effect on modern ice. There is a slight tendency to favor small TAU (less than 0.2, corresponding to less than 2 kyr), compared to the nominal prior value of 3 kyr \citep{pollard2012description}, but this is a weak tendency. 

There is one strong signal in Figure \ref{fig:BinaryIceResultsObs}: the best-fit basal sliding coefficient CRH for continental-shelf (modern ocean) regions, is found to be quite high, 0.8, corresponding to 10$^{-6}$ m a$^{-1}$ P a$^{-2}$. 
The narrowing of the uncertainty for CRH is valuable information for models. Even though the validation here is based only on results for the modern ice sheet (for which there is no direct effect of CRH if the results are realistic, because all modern ice should be floating in the affected regions), CRH still has a significant influence on the modern state by affecting ice profiles during the recent evolution over the last few thousand years. The relatively high (slippery) value of $10^{-6}$ is consistent with geologic data at LGM, indicating that grounded ice thicknesses across the modern Ross Sea Embayment were relatively thin at LGM \citep[deduced from geologic data in the Transantarctic Mountains and other outcrops;][]{ackert2013controls,anderson2013ross}.

In the ice-volume change projections for 500 and 1000 years from present (Figure \ref{fig:IceVolumeProjectionObs}), the left-hand tail of lower ice volumes may seem counter-intuitive, given the imposed climatic warming after year 0 (present day). This tail is probably due to a subset of highly unrealistic runs, whose grounding lines retreat drastically inland beyond the modern coastline much too early, 100's to 1000's years before present, and then re-advance due to the lagged bedrock rebound (discussed above), producing modern grounding lines close to observed despite very unrealistic past behavior. This likely happens in runs with the longest TAU time-scale values, 7 kyrs. As mentioned above, calibration based on  past geologic data would reduce this tail.

\begin{figure}
\centering
\includegraphics[scale=0.4]{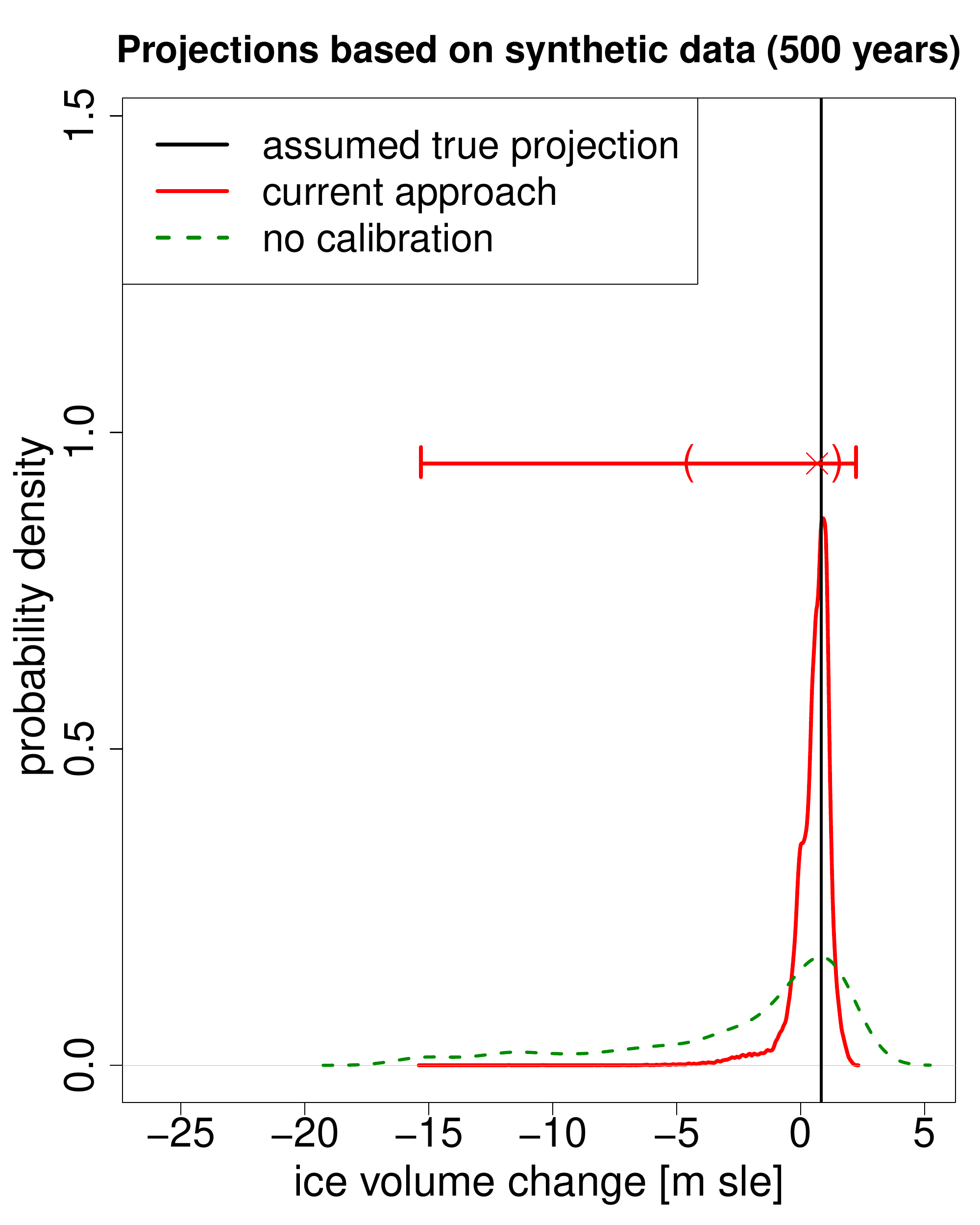}
\includegraphics[scale=0.4]{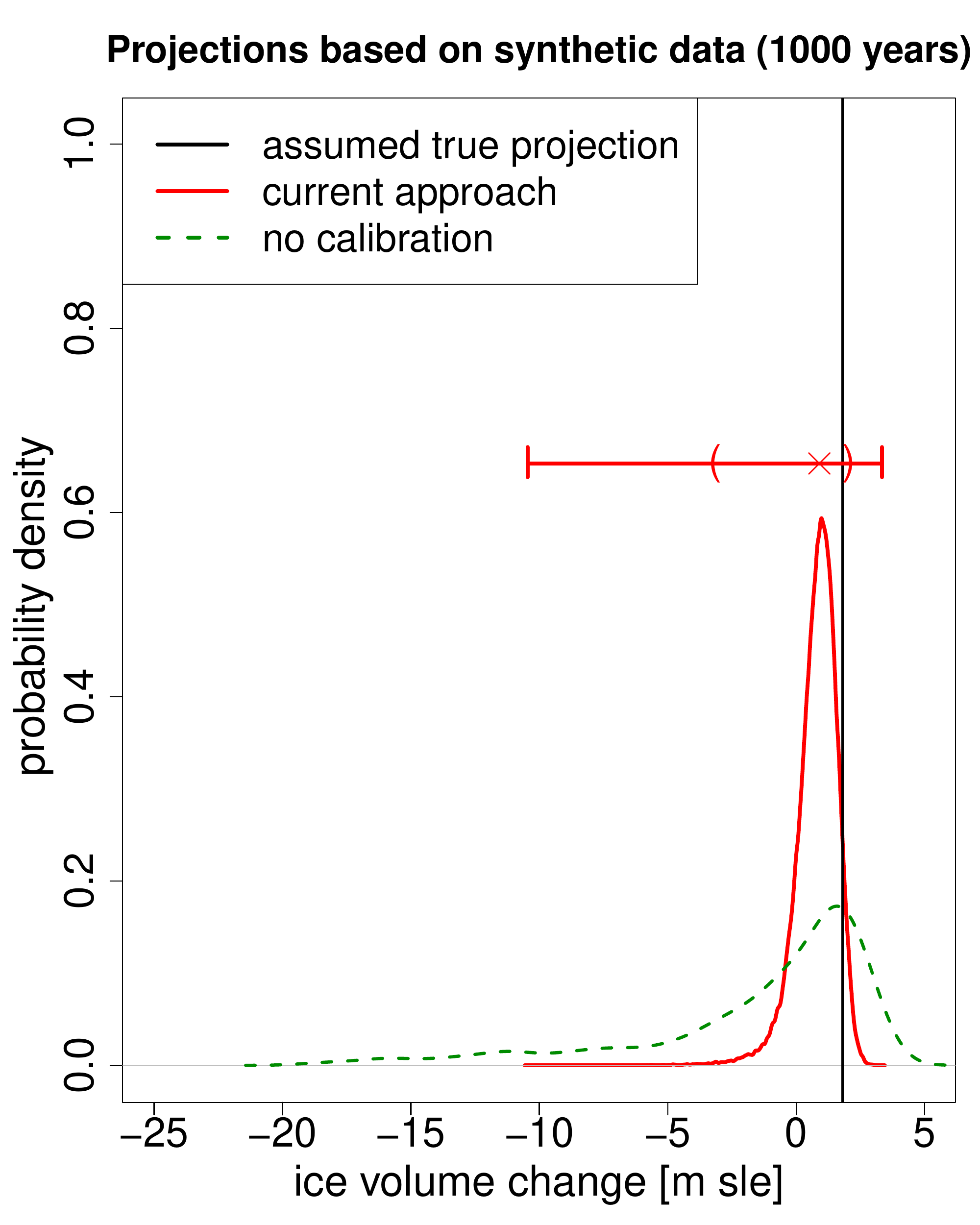}
\caption{The ice volume change projections for 500 years (left panel) and 1000 years (right panel) from the simulated example described in Section \ref{section:CalibrationIce} based on the calibration results (solid red lines) in Figure \ref{fig:BinaryIceResults} and no calibration (dashed green lines). The vertical black line shows the true projection value generated by the synthetic truth. The parentheses on the horizontal bars represent 95\% prediction intervals and the vertical bars the ranges. The slight bias in the  mode of the predictive distribution for the volume changes for 1000 years is likely due to the bias for the posterior mode of TAU (see Figure \ref{fig:BinaryIceResults}).}
\label{fig:IceVolumeProjection}
\end{figure}

The difference between the mode at 1000 years from present and the assumed-true value (Figure \ref{fig:IceVolumeProjection}) could well be due to the unconstrained TAU values in Figure \ref{fig:BinaryIceResults} and \ref{fig:BinaryIceResultsObs}. As well as affecting past behavior, TAU also has an effect on long-term future retreat via the feedback on grounding-line depths discussed above. Nevertheless, the basic projection of drastic future retreat and a few meters (less than 10 meters) equivalent sea-level rise within the next 1000 years is a robust result, and occurs in nearly all runs with reasonably realistic past behavior. This is consistent with other recent studies of future ice-sheet retreat in the Amundsen Sea Embayment sector \citep{joughin2014marine,rignot2014widespread}.

\bigskip
\textit{Caveats.}
Similar to other calibration frameworks, care is needed in specifying the model-observation discrepancy to avoid identifiability issues. In our calibration approach specifying the discrepancy model requires a pre-determined tuning parameter $c$, which determines whether a specific pixel exhibits a persistent model-observation discrepancy (see Section \ref{section:BinaryCalibCalibration}). While our choice of $c=0.5$ appears to be somewhat ad-hoc, we have verified through a detailed study of multiple simulated examples that this choice of $c$ generally yields a discrepancy pattern that captures well the true discrepancy pattern. In addition to the simulated example (Section \ref{subsection:BinarySimulatedExample}) and the PSU3D-ICE model used in this paper, we have applied our approach with $c$ set to 0.5 in the context of a completely different ice sheet model \citep[SICOPOLIS;][]{greve1997application,greve2011initial} and verified that the identified discrepancy is similar to the true discrepancy pattern used to create the simulated examples.

Our scientific conclusions are subject to the following usual caveats. It should be emphasized that the extensions of our runs beyond present are projections based only on a single, simplified greenhouse-gas scenario. Various greenhouse-gas scenarios for the future are available as Representative Concentration Pathways \citep[RCPs][]{moss2010next,meinshausen2011rcp}, which can be used with climate models to provide forcing for the ice sheet model. However, the RCPs themselves range over a wide envelope (RCP 2.6, RCP 4.5, RCP 6.0, RCP 8.5) depending on future societal activity, from strong through intermediate mitigation to ``business-as-usual". These scenarios should ideally also be treated as sources of additional projection uncertainty.   This is a complex additional challenge and a potentially important direction for future work. In this preliminary study we  limit the variable  parameters to  physical model attributes, and specify a relatively simple future climate warming. Note also that the volumes in Figures \ref{fig:IceVolumeProjection} and \ref{fig:IceVolumeProjectionObs} are only for the WAIS domain of these nested model runs, not for all Antarctica \citep[although the East Antarctic contribution is expected to be comparatively small;][]{pollard2009modelling}.

 Our calibration results are based only on the modern day grounding-line geometry and it might be possible to improve the results by using additional information such as the past grounding line positions. We expect that this will lead to a better constrained probability density of TAU. As we mentioned above, one future direction is to formulate a rigorous calibration approach that enables us to combine these multiple sources of information. In addition, we use only the grounding-line geometry and hence ignore the spatial patterns of floating ice (ice shelves) and the distribution of ice thickness. Incorporating these additional sources of information requires calibration using other types of non-Gaussian data such as bounded spatial data with many zeros and spatial data with multinomial responses; this is beyond the scope of this paper. Also, methods combining calibration with other types of past geologic data will be challenging but necessary for comprehensive model validation \citep[mainly Relative Sea Level records, GPS uplift rates, and cosmogenic age-elevation data;][]{briggs2013how, briggs2013glacial,briggs2014data,whitehouse2012deglacial,whitehouse2012new}.

\section{Summary} \label{section:BinaryCalibDiscussion}

We have formulated a novel computer model emulation and calibration approach for high-dimensional binary spatial data. We have applied our approach to calibrate a computer model that describes the dynamics of the West Antarctic ice sheet. The results of this calibration provide some understanding of the role of important parameters in the model, and also allow us to produce  preliminary projections, with uncertainties, of future ice sheet volume change. Using a logistic principal component analysis and basis representation, we can efficiently handle high-dimensional binary spatial data; our dimension-reduced approach therefore addresses challenges in handling high-dimensional binary spatial data by removing the need to simulate underlying high-dimensional processes and avoiding computations with large matrices. The logistic principal component analysis is a special case of principal component analysis for the one parameter exponential family and can  therefore be easily extended to other types of spatial data such as spatial count processes. Incorporating complicated model-observation discrepancies is a challenging problem in general. We allow for a flexible but computationally expedient model for discrepancy. Our study of simulated examples shows that our approach can estimate the input parameters reasonably well even in the presence of considerable, realistic model-observation discrepancies. The application of our methods to real ice sheet observational data allows us to characterize uncertainties in both our model parameter estimates as well as our projections for future ice volume change. 

\subsection*{Acknowledgments}
We are grateful to Zhengyu Liu and his group at U. Wisconsin for providing output from their coupled GCM simulation \citep{liu2009transient}, used here to prescribe ocean temperatures over the last 20,000 years. We also thank A. Landgraf for distributing his code for logistic PCA freely on the Web (https://github.com/andland/SparseLogisticPCA).

This work was partially supported by National Science Foundation through (1) NSF-DMS-1418090 and (2) NSF-OPP-ANT-1043018, (3) Network for Sustainable Climate Risk Management (SCRiM) under NSF cooperative agreement GEO–1240507, and (4) NSF Statistical Methods in the Atmospheric Sciences Network (Award Nos. 1106862, 1106974, and 1107046). WC was partially supported by (3) and (4), and MH and PA were partially supported by (1) and (3), and DP is partially supported by (1), (2) and (3).  All views, errors, and opinions are solely those of the authors.

\bibliographystyle{plainnat}
\bibliography{../long,../references}

\clearpage

 \begin{table}[hc]
\centering
\begin{tabularx}{\textwidth}{ |l|X|}
\hline
Parameter & Definition \\
\hline
OCFAC&  Ocean sub-ice-shelf melt factor (.1 to 10) (non-dim).
         K in Eq. 17, PD12.\\

 CALV & Calving factor (.316 to 3.16) (non-dim), multiplies combined crevasse depth to thickness ratio r in Eq. B7, PD15.\\

 CRH  &  Basal sliding coeff. modern ocean bed (10$^{-9}$ to 10$^{-5}$) (m yr$^{-1}$ P a$^{-2}$)
         C in  Eq. 11, PD12.\\

 CRHASE &  Extra multiplicative factor$\times$CRH, inshore ASE (10$^{-3}$ to 1) (non dim),
         multiplies C, only in inshore Amundsen Sea Embayment, Eq. 11, PD12.\\

TAU  &   Asthenospheric relaxation \textit{e}-folding time scale= (1 to 6 kyr).
        $\tau$ in Eq. 33, PD12.\\

 LITH  &  Lithospheric stiffness (10$^{23}$ to 10$^{25}$ ) (N m).
         D in Eq. 30, PD12.\\

 GEO &   Geothermal heat flux (50 to 90) (mW m$^{-2}$).
         see section 2.7, PD12.\\

 SUBPIN & Sub-grid ice-shelf pinning factor (0 to 4) (non-dim),
         multiplies $s_{dev}$ in Eq. 13, PD12.\\

 USCH  &  Grounding-line flux factor (0.5 to 5) (non-dim),
         multiplies $q_g$ in Eq. 8, PD12.\\

 LAPSE  & $\gamma$ (lower-case greek) in Eq. 34a, PD12.
         atmospheric lapse rate (-.005 to -.010) ($^{\circ}$C m$^{-1}$).\\

\hline
\end{tabularx}
\caption{ Ice-sheet model parameters varied in the LHC ensemble described in Section \ref{section:DataDescription}. The parameter ranges and units are shown in parentheses. The parameter values are uniformly spaced within each range, but for some variables, they are log-linear (i.e., the log of the parameter is varied uniformly), for OCFAC, CALV, CRH, CRHASE, LITH, and USCH.
Parameters called "factors" are non-dimensional, multiplying existing 
terms. The corresponding symbols and equation numbers in \citet{pollard2012description} and \citet{pollard2015potential} called PD12 and PD15 respectively, are also shown in the table.}
\label{table:parameters}
\end{table}

\end{document}